\documentclass[tighten]{aastex6}
\usepackage{amsmath,amssymb,amsthm}
\usepackage{natbib}
\usepackage{graphicx}
\usepackage{CJKutf8}

\slugcomment{}

\shortauthors{Zheng, Lin \& Kouwenhoven}

\begin{document}
\begin{CJK*}{UTF8}{gbsn}
\title{Planetesimal clearing and size-dependent asteroid retention by
secular resonance sweeping during the depletion of the solar nebula}

\author{Xiaochen Zheng(郑晓晨)\altaffilmark{1,2,3,4,8}}
\author{Douglas N.C. Lin(林潮)\altaffilmark{2,3,5,6,9}}
\author{M.B.N. Kouwenhoven(柯文采)\altaffilmark{7,2,10}}
\altaffiltext{1}{Department of Astronomy, Peking University, Yiheyuanlu~5, Haidian District,
Beijing 100871, P.R. China}
\altaffiltext{2}{Kavli Institute for Astronomy and Astrophysics, Peking University, Yiheyuanlu~5, Haidian District,
Beijing 100871, P.R. China}
\altaffiltext{3}{Department of Astronomy and Astrophysics, University 
of California, Santa Cruz, CA 95064, USA}
\altaffiltext{4}{Center for Astrophysics, Tsinghua University, 
Shuangqing Rd.~30, Haidian District, Beijing 100084, China}
\altaffiltext{5}{Institute for Advanced Studies, Tsinghua University, 
Shuangqing Rd.~30, Haidian District, Beijing 100084, China}
\altaffiltext{6}{National Astronomical Observatory of China, Datun Rd., 
Chaoyang District, Beijing 100012, P.R. China}
\altaffiltext{7}{Department of Mathematical Sciences, Xi'an Jiaotong-Liverpool University, 111 Ren'ai Rd., Suzhou Dushu Lake Science and Education Innovation District, Suzhou Industrial Park, Suzhou 215123, P.R. China}
\altaffiltext{8}{x.c.zheng1989@gmail.com}
\altaffiltext{9}{lin@ucolick.org}
\altaffiltext{10}{t.kouwenhoven@xjtlu.edu.cn}

\begin{abstract}
The distribution of heavy elements is anomalously low in the asteroid 
main belt region compared with elsewhere in the solar system. 
Observational surveys also indicate a deficit in the number of small 
($ \la 50$~km size) asteroids that is two orders of magnitude lower than what 
is expected from the single power-law distribution that results from a
collisional coagulation and fragmentation equilibrium.  Here, we consider 
the possibility that a major fraction of the original asteroid population 
may have been cleared out by Jupiter's secular resonance, as it swept through 
the main asteroid belt during the depletion of the solar nebula. This effect 
leads to the excitation of the asteroids' orbital eccentricities. 
Concurrently, hydrodynamic drag and planet-disk tidal interaction 
effectively  damp the eccentricities of sub-100 km-size and of 
super-lunar-size planetesimals, respectively. 
These combined effects lead to the asteroids' orbital decay and clearing 
from the present-day main belt region ($\sim 2.1-3.3$~AU).  Eccentricity 
damping for the intermediate-size (50 to several hundreds of kilometers) 
planetesimals is less efficient than for small or large planetesimals. 
These objects therefore preferentially remain as main belt asteroids near their birthplaces, with modest asymptotic eccentricities.  The smaller asteroids are the 
fragments of subsequent disruptive collisions at later times as suggested 
by the present-day asteroid families.  This scenario provides a natural explanation 
for both the observed low surface density and the size distribution of 
asteroids in the main belt, without the need to invoke special planetesimal 
formation mechanisms.  It also offers an explanation for the confined spatial 
extent of the terrestrial planet building blocks without the requirement of 
extensive migration of Jupiter, which is required in the grand-tack scenario.  
\end{abstract}


\section{Introduction}
The formation and evolution of planetesimals are essential steps 
in the classical core accretion model for the origin of planets
\citep{pollack1996, idalin2004}.  The conventional coagulation
scenario \citep{safronov1969, wetherill1980} is based on the 
assumption that these planetary building blocks grow due to
cohesive collisions through a runaway process, followed by an oligarchic 
phase \citep{kokuboida1998}.  The crater-covered surfaces of 
asteroids, the Moon, and Mercury provide vivid supporting 
evidence for such an assumption.  These collisions are generally
preceded by close elastic encounters that excite the 
planetesimals' eccentricities and induce orbit crossing.

Through scattering and collisions, planetesimals attain an 
equilibrium velocity dispersion, $\sigma$. In a gas-free environment,
the magnitude of $\sigma$ is a significant fraction of their 
characteristic surface escape speeds \citep{aarseth1993,palmer1993,  
kokuboida1998}.  At such a high speed, many collisions among 
super-kilometer-size planetesimals may lead to breakup rather 
than a merger \citep{agnor2004, leinhardt2012, stewart2012}. 
Direct evidence of collisional fragmentation can be found 
in iron/stone meteorites. Their parent bodies were 
differentiated prior to catastrophic collisions. The 
much more common chondritic meteorites may also be the 
collisional by-products of parent bodies that may have 
avoided differentiation because they formed after the 
radioactive $^{26}$Al isotopes have mostly decayed.

The possibility of collisional fragmentation poses a
potential barrier for planetesimal formation and growth.
In attempts to explore pathways to bypass the kilometer-size coagulation 
barrier, several scenarios have been proposed. Grains and 
planetary building block materials may be trapped in regions
with a local pressure maximum, where the flow is Keplerian
and the planetesimals' $\sigma$ is relatively small, to promote 
cohesive rather than disruptive collisions.  One such a 
location is the snow line \citep{kretke2007,brauer2008}
which, during the advanced stages of solar nebula evolution,
may have been located close to or interior to the present-day asteroid belt
\citep{garaudlin2007}. Other growth hypotheses that have been proposed 
and analyzed so far include gravitational instability 
\citep{goldreichward1973, weidenschillingcuzzi1993, youdinshu2002, 
garaudlin2004}, streaming instability \citep{youdingoodman2005}, 
and turbulent trapping \citep{cuzzi1993, johansen2007}.

While most planetesimals may eventually be accreted by a 
few massive embryos to form either terrestrial planets 
or progenitor cores of gas giants \citep{idalin2004}, some 
relics are retained in the asteroid belt. The dynamics, 
structure, and composition of the asteroids carry important 
information on the chronology and dominant physical processes 
associated with planetesimal and planet formation.  One particularly 
important clue is the asteroids' size-frequency distribution 
(SFD).  Conventional coagulation models predict a power-law 
spectrum \citep{dohnanyi1969} that generally matches that 
of impactors thought to have produced the observed crater 
size distributions on the Moon and on Mercury.  In contrast, 
the alternative collective mechanisms lead to a rapid emergence 
of preferentially large planetesimals.

The observed SFD in the present-day main belt ($\sim 2.1-3.3$ AU) 
is dominated by the midsize asteroids, with an apparent lack of small 
(sub-kilometer-size) and large (moon-size) bodies \citep{bottke2005}.  
Based on the assumption that planetesimals formed through 
collective mechanisms \citep{johansen2007, cuzzi2008} with 
a minimal size around 100~km, \cite{morbidelli2009}
reproduced the observed SFD slope.  However, the rapid 
formation of relatively large planetesimals also implies 
that they are likely to have acquired similar quantities 
of radioactive $^{26}$Al isotopes, as those found in the most 
primitive Calcium-Aluminium-rich inclusions (CAIs). Under such 
conditions, the heat released from nuclear fission would be adequate 
to melt and differentiate planetesimals with sizes larger than a 
few tens of kilometers \citep{mcsween1999}. Under the rapid formation scenario, 
first-generation large planetesimals are likely to be differentiated.
Cosmochemical analysis of iron meteorites \citep{kelley2000}
suggests that there may indeed have been $50-100$ differentiated parent bodies 
that acquired their relatively large mass at sufficiently early time.
However, the undifferentiated chondritic meteorites make up the 
predominant population of meteorites that struck the Earth. 
Since most meteorites originate from the asteroid belt, if a 
population of large planetesimals did emerge very early (within
a few hundred thousand years), they and their collisional 
fragments would have to be preferentially cleared out of the 
main belt region.

The empirical \emph{minimum mass solar nebula} (MMSN) model 
\citep{hayashi1981, weidenschilling1997} and the conventional 
formation model for large asteroids and chondritic meteorites 
in the main belt are constructed based on the assumption that the 
mass distribution (in both gas and refractory solids) is continuous 
throughout the solar nebula \citep{wetherill1989, connolly1998, 
ciesla2002, desch2002, johansen2007}. However, the total
present-day mass of the asteroids, estimated from Mars's orbit 
and the asteroids' observed SFD, is $\sim 6 \times 10^{-4}$ 
Earth masses \citep{morbidelli2009}. This apparent depression in the mass
distribution, relative to the MMSN
model, suggests that up to $99.9\%$ of the residual planetesimals'
total mass may have been lost from the main belt region. In the 
context of the ``grand-tack'' model, \citet{walsh2011} put forward 
the possibility of inward-then-outward migration of Jupiter and Saturn, 
which may have migrated to the present-day location of the main 
belt prior to the disk depletion. As a consequence, asteroids in the 
main belt are severely depleted and repopulated during the phase 
of the gas giants' instability.

In this work we propose an alternative scenario to account for 
the observed SFD and the mass deficit in the main asteroid belt, 
based on the classical planetesimal coagulation (rather than 
the collective formation) model.  We assume that coagulation 
and fragmentation 
of asteroids occurred within 2~Myr (which is comparable 
to the radiogenic age of the chondritic meteorites)
when the energy release rate from the decay of $^{26}$Al has 
largely diminished.  We also assume that these processes lead 
to a continuous SFD ranging from dust to lunar-size protoplanetary 
embryos \citep{chambers2008} with a total mass several times that 
of the Earth.

In order to reproduce the observed SFD and the present-day low surface 
density in the main belt region, we propose that the small 
and large planetesimals were cleared by $\nu_{5,6}$ secular 
resonance (SR) that swept through the region during the local 
clearing and global depletion of the solar nebula
\citep[e.g.][and references therein]{heppenheimer1980, ward1981, lemaitre1991, lecar1997,nagasawaida2000, nagasawa2000, nagasawa2001, nagasawa2002, obrien2007}. We suggest that this process occurred after the formation of Jupiter and on a time scale
 of $\sim 3-5$~Myr, comparable to the observed depletion time 
scale of protostellar disks \citep{hartmann1998} and  
the radiogenic age difference between the CAIs and chondrules.

The sweeping secular resonances (SSRs) excite the eccentricities of 
all planetesimals along their path.  The eccentricities of the 
small (sub-kilometer-size) rocky bodies and large (Moon-size) embryos 
are effectively damped by the hydrodynamic drag and planet-disk 
tidal interaction.  Consequently, these planetesimals undergo 
orbital decay synchronously with the inward sweeping of the 
$\nu_5$ SR \citep{nagasawa2005, thommes2008}.  As an extention
of these previous investigations, we show that some intermediate-size planetesimals are retained in the main belt because their
eccentricity damping and orbital decay are less effective.  
 
This paper is structured as follows.  In \S2, we briefly 
recapitulate the basic physical effects of the sweeping SRs, and we describe our numerical method and initial conditions in \S3. 
In \S4 we compute the orbital evolution for 
planetesimals with a range of sizes. Based on these results, 
we reconstruct the asteroids' SFD and the mass depletion 
rate of the asteroid belt under the combined effects of 
SSRs and eccentricity damping.  We 
compare our results with observations and discuss the 
implications in \S5.


\section{Brief description of the SSRs}
\label{sec:quantitive}

We briefly recapitulate the physical concept of the dynamical 
shake-up model, which was proposed in the context of terrestrial 
planet formation by \cite{nagasawa2005} and \cite{thommes2008}. This model is based on the following assumptions that (i) the initial growth 
of planetesimals of an unperturbed gaseous solar nebula was 
limited by their isolation mass $M_{\rm iso}$.  In an MMSN, $M_{\rm iso} < M_\oplus$ within a few astronomical units, but beyond the snow line embryos may acquire super-Earth masses
\citep{idalin2004}.  (ii) Relatively massive ($> 10\,M_\oplus$) 
embryos can efficiently accrete gas \citep{pollack1996} and evolve 
into Jupiter and Saturn within $\sim 2$\,Myr. (iii) Gas in the solar 
nebula was depleted over a characteristic time scale ($\sim 3-5$\,Myr), comparable to that observed in disks around T~Tauri stars
\citep{hartmann1998}.

In addition, we assume that gravitational interaction between the 
emerging gas giants and nearby embryos and planetesimals induced 
scattering and giant impacts during their formation \citep{
liagnorlin2010, idalin2013}. These close encounters lead not only 
to orbit crossing \citep{zhou2007} and compositional 
mixing of residual planetesimals \citep{demeo2014}, but 
also to eccentricity excitation of the emerging planets. 
With a finite eccentricity, Jupiter and Saturn exert a secular 
perturbation on the residual planetesimals, causing their eccentricities 
to modulate and their longitudes of periastron to precess \citep{murray1999}. 
This perturbation is particularly strong near the gas giants'
low-order mean motion resonances (MMRs).  Due to the self-gravity 
of the nebula, Jupiter's and Saturn's orbits also precess. In regions 
where two precession frequencies match, the SRs 
excite the planetesimals' eccentricities as angular 
momentum is monotonically transferred from the planetesimals to 
the gas giants.

In principle, all the planets contribute to the secular perturbation. 
Two particularly strong SRs among these are the $\nu_5$ 
and $\nu_6$ SRs.  They are dominated by the perturbations from 
Jupiter and Saturn, respectively \citep{agnorlin2012}. In this paper,
we include these additional planetary contributions.  However, 
contribution to the gas giants' precession rates by the disk potential 
is comparable to that induced by Jupiter's and Saturn's interactions.
When the solar nebula's mass distribution was comparable to that of the 
MMSN model, the $\nu_5$ SR was 
located near the asteroids' main belt. During the advanced stages of nebula 
evolution, the precession rates induced by the disk decline with its 
diminishing surface density ($\Sigma$), resulting in the locations of 
these resonances sweeping through the solar system.  
Today, the nebula is completely cleared of gas and the $\nu_5$ SR is located interior to the 
orbit of Venus, whereas the $\nu_6$ SR is beside the terrestrial 
planet region, between the previous orbit of Mars and the main asteroid belt.

\section{Computational models}
\label{sec:computation}

Our work mainly focuses on the combined 
effects of two competitive processes: eccentricity excitation by 
MMRs and SRs and eccentricity damping of the progenitors of 
the modern-day asteroids during the advanced stages of solar 
nebula evolution (a few Myr after CAI formation). In this 
section we show that both small planetesimals and large embryos are 
preferentially cleared out from the main asteroid belt region, and discuss
how these processes result in the observed SFD of the present-day asteroid belt.

\subsection{Numerical method}
\label{sec:nmethod}

In this paper we propose that orbital decay associated with
the sweeping SR is the main cause of size-selected 
clearing of the asteroid population from the main belt region.  
In order to quantitatively establish this conjecture, we present 
the results of a series of numerical simulations. The computational 
tool we employ is a modified version of the publicly available HERMIT4 package  \citep{aarseth2003}, which is ideally suited to carry out planetary system simulations.

We consider systems consisting of the Sun, Jupiter, and Saturn. For most
models, including the default model, the two gas giant planets are 
initialized at their present-day locations, with eccentricities slightly 
different from their present-day values. The 
planetesimals are represented by a population of coplanar (with 
respect to the giant planet) particles with initial semimajor 
axes distributed in the current domain of the main asteroid belt. 
These particles can be treated as massless, such that 
we can neglect their gravitational perturbation on the disk and on the gas 
giants, as well as their mutual gravitational interactions.  

The internal density of asteroids inferred from their orbital 
dynamics is in the range $1-5~\rm{g/cm^3}$ \citep{margot2002,
marchis2006, descamps2008}. We assume that differentiation and fragmentation 
may have occurred among relatively large planetesimals, whereas small
planetesimals may be considered to be pristine rubble piles.  We 
construct a simple prescription for the planetesimals' 
internal density, ranging from that of iron/stone meteorites with 
zero porosity ($\rho_{p} = 5.5~\rm{g/cm^3}$) for the relatively large 
(with radius $r_p > 100$~km) planetesimals to that of water ice 
($\rho_p = 1~\rm{g/cm^3}$) for the small planetesimals ($r_p < 18$~km).  We also 
assume a transitional density $\rho_p(r_p) = (r_p/18~{\rm km})$~g/cm$^3$ 
for the intermediate-size (18~km $ \le r_p \le$ 100~km) planetesimals.

For the default model, we assume that both Jupiter and Saturn have 
already obtained their present-day masses. During the passage 
of $\nu_5$ and $\nu_6$, planetesimal eccentricity 
excitation occurs due to angular momentum transfer from these smaller 
bodies to Jupiter and Saturn (without any energy exchange). The 
strength of the torque is a function of Jupiter's and Saturn's 
eccentricities ($e_J$ and $e_S$, respectively).  These quantities also
oscillate with an amplitude that is determined by the total angular 
momentum deficit of the system, i.e., the difference between the actual 
sum of Jupiter's and Saturn's angular momentum and the maximum values
for their given semimajor axes ($a_J$ and $a_S$, respectively).  

Since Jupiter's mass is substantially larger than that of Saturn, 
it is adequate to specify the angular momentum deficit of the system by 
assigning an initial non-zero eccentricity, $e_{J}$ to Jupiter only. 
While Saturn is initialized with a circular orbit, its eccentricity
is rapidly excited by Jupiter.
In principle, we can adopt the initial value of $e_J$ from the current
angular momentum deficit of the solar system, or from the results of other
numerical simulations.  However, the angular momentum deficit of the 
system may have evolved due to the gas giants' interaction with residual
planetesimals and gas. Taking these uncertainties into account, we consider two values for Jupiter's initial eccentricity, which are slightly larger and slightly smaller than Jupiter's present-day value, respectively. In the next section we discuss the consequences of these choices for $e_J$.

\subsection{Contribution of gravity and damping by the gas disk}

We mainly focus on the advanced stage of nebula evolution when 
the giant planets' gas accretion is quenched by gap formation
and type II migration is stalled by the depletion of the disk
gas \citep{lin1986, dobbs2007}. In our model we assume (i) a thin-disk 
approximation, (ii) an asymmetric power-law surface density distribution 
based on the MMSN model \citep{hayashi1981}, 
with (iii)  an axisymmetric gap near the gas giant \citep{bryden1999},  and  (iv) that all planetary components (gas giants and planetesimals) are coplanar in the protoplanetary disk.

We adopt a thin-disk model with an axisymmetric surface density distribution, $\Sigma(r,t)$, of the form
\begin{equation}
\Sigma(r,t) = \Sigma(r_0,0) \, f_{\rm dep}(t) \, (r/r_{0})^{-k} \quad ,
\label{eq:sigma}
\end{equation}
where $r_0=1$~AU and $f_{\rm dep}(t) = {\rm exp} (- t/T_{\rm dep})$ is the 
time-dependent depletion function of the gas disk. In the MMSN model, $k = 1.5$ is widely used with an initial surface density $\Sigma(r_0,0) = 1700~\rm{g/cm^{2}}$ at the location of $r_0=1$~AU.
We approximate the density of the residual gas
to be $\rho_{g} = \Sigma (r, t) / H (r)$, with a disk scale height 
that has a similar form to that in the work of \cite{hayashi1981} 
and \cite{thommes2008}, but is reduced by a factor of two: 
\begin{equation}
H(r) = 0.025 \left(\frac{r}{1~{\rm AU}}\right)^{5/4}\ {\rm AU} \quad 
\end{equation}
(see further discussions in \S\ref{sec:disk}).
This relatively small thickness is appropriate for the advanced stages
of disk depletion \citep{garaudlin2007}.

The characteristic decay time scale is chosen to be $T_{\rm dep}= 1$~Myr 
for the default models. The initial time ($t = 0$) is set to be the start 
of gas depletion (rather than the epoch of CAI formation). This definition 
is consistent with the assumed prior emergence of Jupiter at the onset of 
our computation.  In comparison with observational data, $T_{\rm dep}$ 
corresponds to the duration for transition from classical to weak-line 
disks. This time scale may be somewhat shorter than the average age of 
stars with detectable IR excess in their continuum spectral energy 
distribution, though it is comparable to the observationally inferred 
evolutionary time scales for transitional disks
\citep{currie2011}.  
Although the detailed dynamical evolution of the 
residual planetesimals may depend on the chosen model parameters, 
including the functional forms of $\Sigma(r,t)$ and $T_{\rm dep}$, 
this working disk model provides an adequate set of initial conditions 
to generate several illustrative examples.  


We compute the disk gravity on the gas giants (Jupiter and Saturn) and the planetesimals separately, because the apsidal precession of the planetesimals by the gas disk is dominated by the gas in their neighborhood, whereas the gas giants' apsidal precession caused by disk gravity is sensitive to the gap structure \citep{nagasawa2005}. Gas giants that have opened a gap experience a 
gravitational force that depends on the locations of the inner and 
outer boundary of the gap \citep{ward1981, nagasawaida2000}: 
\begin{equation}
F(r,t) = 2\pi G \Sigma(r,t) \sum_{n=0}^{\infty} A_{n} 
		\left\{ 
		\left( \frac{2n}         {2n-1+k}       \right)
		\left( \frac{r}          {R_{\rm out}}  \right)^{2n-1+k}  	 
	  -\left( \frac{2n+1}       {2n+2-k}       \right)
  	    \left( \frac{R_{\rm in}} {r}            \right)^{2n+2-k} 
	 	\right\} \  .
\end{equation}
Several models are presented here.  In all models, we set the semimajor axis 
of Jupiter to its present-day value (5.2~AU) and  $R_{\rm in} = 4.5~$AU 
in accordance with the results of numerical simulations \citep{bryden1999}.
In the default model and several other models, we also set Saturn's semimajor axis
to its present-day value (9.58~AU) and adopt $R_{\rm out} = 11.0~$AU following
the results of numerical simulations \citep{bryden2000}.  In all models
we find that the dynamical evolution of Jupiter's and Saturn's orbits and the 
propagation of the $\nu_5$ and $\nu_6$ SRs depend sensitively on $R_{\rm in}$ and weakly on $R_{\rm out}$.  For 
computational convenience, we adopt the same values of $R_{\rm in}$ and 
$R_{\rm out}$ in all models presented here.

The effect of gravity on the planetesimals is more complex. Even though 
the gravity experienced by the planetesimals is usually dominated by 
bodies in the nearby region, planetesimals in the region close to the inner edge 
($R_{\rm in}$) of the truncated disk are also affected by the presence 
of the gap. We therefore combine the prescription constructed in 
\cite{nagasawa2005} with the ``gap effect'' of \cite{ward1981} to 
compute the disk's gravity on the planetesimals located at radius $r$.  
We find
\begin{equation}
F = - 4\pi G \Sigma(r,t)
	\left\{
	Z_k + \sum_{n=0}^{\infty} A_{n} 
	\left(\frac{n}{2n-1+k}\right)
	\left(\frac{r}{R_{\rm in}}\right)^{2n-1+k}
	\right\}.
\label {eq:f_pd}
\end{equation}
%

While MMRs and SRs excite the orbits of planetesimals, 
residual gas in the nebula also damps the eccentricity 
of the planetesimals. For small 
(sub-kilometer-size) planetesimals, this process operates mainly through
hydrodynamic drag \citep{adachi1976}. For large (super-moon-size) 
embryos, the main eccentricity-damping mechanism is planet-disk 
tidal interaction \citep{goldreichtremaine1980, ward1997}.  
Whereas the SRs remove angular momentum and preserve 
the energy of planetesimal orbits, eccentricity damping 
leads to the dissipation of energy without changes in the 
angular momentum of their orbits.  Consequently, planetesimals 
undergo inward migration during orbital circularization.
For those planetesimals whose eccentricities are damped efficiently, 
the orbital decay is kept in pace with the inward propagation 
of the $\nu_5$ SR.  These planetesimals ``surf'' 
over extended radial distances.

With a simple approximation \citep{zhou2007} 
we compute the effect of eccentricity damping with a drag force
that reduces the amplitude of the planetesimals' epicyclic motion
around their guiding centers such that
\begin{equation}
{\bf F}_{\rm D} = - \frac{{\bf V} - {\bf V}_{\rm kep}^{\prime}}
{T_{\rm damp}} \quad .
\label{eq:dragforce}
\end{equation}
In this expression, the motion of the gas is assumed to be axisymmetric 
and in the azimuthal direction with an amplitude $V_{\rm kep}^{\prime} 
= V_{\rm kep}(1-\eta)$, where the dimensionless parameter 
$\eta \approx (c_s/4V_{\rm kep})^2$, with $c_s$ as the sound 
speed, represents the contribution from the
pressure gradient \citep{adachi1976}.

The magnitude of $T_{\rm damp}$ is different for the small and large
planetesimals.  The eccentricities of small planetesimals are mainly damped by hydrodynamic drag. Since the planetesimal radii $r_p$  are much larger than 
the molecular mean free path in the gas, the Stokes drag law 
\citep{whipple1972, adachi1976, weidenschilling1977, ida1996, 
supulver2000} can be applied in our model, and it operates on a timescale
\begin{eqnarray} 
T_{\rm damp,s} = \frac{1}{C_D} \frac{\rho_{p}}{\rho_{g}} 
\frac{r_{p}}{ V_{\rm rel}} 
\simeq {43 \over f_{\rm dep}}
             \left(\frac{\rho_{p}} {1~{\rm g}/{\rm cm}^{3}}\right) 
             \left(\frac{r_p} {1~{\rm km}} \right) 
             \left(\frac{a} {1~{\rm AU}}\right)^{11/4} 
             \left(\frac{V_{\rm rel}}{1~{\rm km/s}}\right)^{-1} 
             {\ \rm yr} \quad  ,
\label{eq:t1}
\end{eqnarray}
where $\rho_p$ is the internal density of the planetesimals, $V_{\rm rel}= 
V - V_{\rm Kep}^{\prime}$ is the relative velocity between gas and asteroids.  
This expression has a factor of 3/8 discrepancy from that in \cite{whipple1972}, 
and in the limit of a large Reynolds number, the drag coefficient is $C_{\rm D} = 0.165$.  

The eccentricity of large embryos is damped by the torque associated with 
their tidal interaction by the disk gas at their Lindblad resonances 
\citep{ward1989, ward1993, artymowicz1993, thommes2008} on a time scale 
\begin{eqnarray} 
T_{\rm damp,t} \simeq 
	\left(\frac{M_{\ast}}{M_{p}}\right)
	\left(\frac{M_{\ast}}{\Sigma a^2} \right)
	\left(\frac{H}{a}\right)^4 \Omega_{k}^{-1} 
	\simeq {3 \times 10^{-4} \over f_{\rm dep}} 
	\left(M_p \over {M_\odot}\right)^{-1}
	\left( {a \over {1~{\rm AU}}} \right)^2 {\rm yr} \quad .
\label{eq:t2}
\end{eqnarray}
In addition, the corotation torque \citep[e.g.,][]{paardekooper2011} 
also contributes to the eccentricity damping. The corotation torque 
has a similar dependence on the surface density and temperature 
distribution in the disk and a magnitude that is a few times
larger than the Lindblad torque.  For simplicity, we artificially 
enhance the total tidal damping from the expression in Equation 
(\ref{eq:t2}) by a factor of five. Note that $T_{\rm damp,t} \propto 
M_p^{-1}$, so that the efficiency of tidal damping increases with 
the mass of the planetesimals and embryos.

As the efficiency of eccentricity damping decreases over time, the locations 
of the $\nu_5$ and $\nu_6$ SRs move inward as the solar nebula is 
globally depleted over several $T_{\rm dep}$.  Nevertheless, the combined 
effect of the mechanisms described above can clear a large fraction of 
the residual planetesimals from the main asteroid belt region. 
After $t > 10 \times T_{\rm dep}$, gas is severely depleted in the disk
and the $\nu_5$ SR passes through the present-day orbit of Mars. The 
concentration of planetesimals around $1-2$~AU increases, and 
as a result, their collisional probability increases. Based on the results of
their numerical simulations, \cite{thommes2008} suggest that 
the associated giant impacts may have promoted the final 
assemblage of the Earth and Mars, as well as the giant impact that
has led to the formation of the Moon. The chronology of 
this scenario is consistent with the 
age estimate of the Earth based on radiogenic dating of Hafnium isotopes \citep{kleine2009}.


\section{Numerical results}
\label{sec:result}

\subsection{SR sweeping}
\label{sec:sweeping}

For our fiducial models we adopt Jupiter's present-day eccentricity 
$e_J = 0.05$ as our initial condition.  The embedded planetesimals
are initialised  with a semimajor axis distribution
\begin{equation}
d N/d a \propto a^{-1.5} \quad \quad (2.0~{\rm AU} \le a \le 3.5~{\rm AU}) \ .
\end{equation}
The contribution of the hydrodynamic drag and the planetesimal-disk tidal interaction
imply that the total damping efficiency is closely correlated with the 
size of the planetesimals. 

\begin{figure*}
\centering
\includegraphics[width=0.95\linewidth,clip=true]{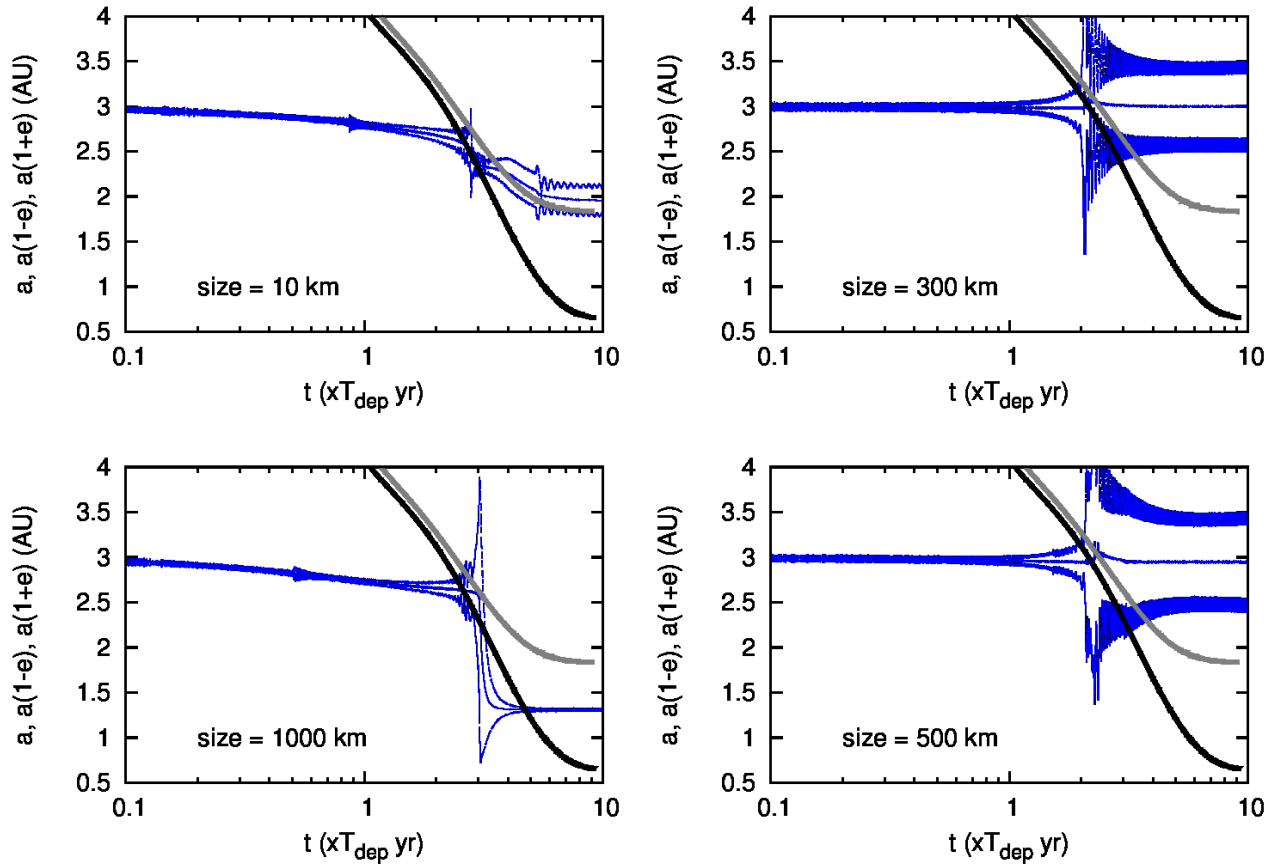}
\caption{Perihelion, semimajor axis and aphelion evolution for four 
representative planetesimals (with $r_p$ = 10, 300, 500, and 1000 km.
In all cases, the initial semimajor axis is $a_p=3$ AU. 
The black and gray curves refer, respectively, to the locations of the $\nu_5$ 
and $\nu_6$ SRs obtained from analytic calculations.}
\label{fig:a_t_ap3}
\end{figure*}

Figure \ref{fig:a_t_ap3} shows how the sweeping SR clears 
planetesimals from a representative primordial location (3~AU).  
We consider four representative cases. The left panels show 
the results for small (10~km) and large (1000~km) planetesimals, 
which are significantly influenced by gas drag and type I damping, 
respectively. The right panels present the results for 
intermediate-size planetesimals (300 and 500~km).  The efficiency 
of eccentricity damping for the intermediate-size planetesimals 
is lower than both the small and the large planetesimals. 
 
At the early stage of evolution, the small and large planetesimals (left
panels in Fig.~\ref{fig:a_t_ap3}) undergo inward migration due to hydrodynamic 
and type I tidal drag, respectively.  In contrast, the intermediate-size planetesimals preserve their initial semimajor 
axes (right panels).  The $\nu_5$ and $\nu_6$ SRs sweep through this
region after $\sim 2T_{\rm dep}-4T_{\rm dep}$ and all the planetesimals experience a 
rapid eccentricity excitation.  Efficient eccentricity damping 
causes the 10 and 1000~km sized planetesimals to undergo considerable 
inward migration. Due to the relatively low eccentricity-damping efficiency, 
a fraction of the planetesimals with intermediate sizes remain near their 
initial locations.  However, most of them are also lost from the main
belt region.  The retention probability is determined by the 
planetesimals' orbital phase during the SRs' passage through their 
semimajor axis. The retained planetesimals have modest residual 
eccentricities.

Based on these results, we infer that (i) nearly all planetesimals larger 
than the Moon or smaller than $\sim 10$~km are likely to undergo extensive 
orbital decay and are evacuated from 3 AU and (ii) some, but not all, 
intermediate-size asteroids may survive the passage of the $\nu_5$ and 
$\nu_6$ SRs and retain their initial semimajor axis at 3 AU.  
These outcomes provide a potential explanation for the depletion and size selection of residual planetesimals from the asteroid main belt region. 

\begin{figure*}
\centering
\includegraphics[width=0.95\linewidth,clip=true]{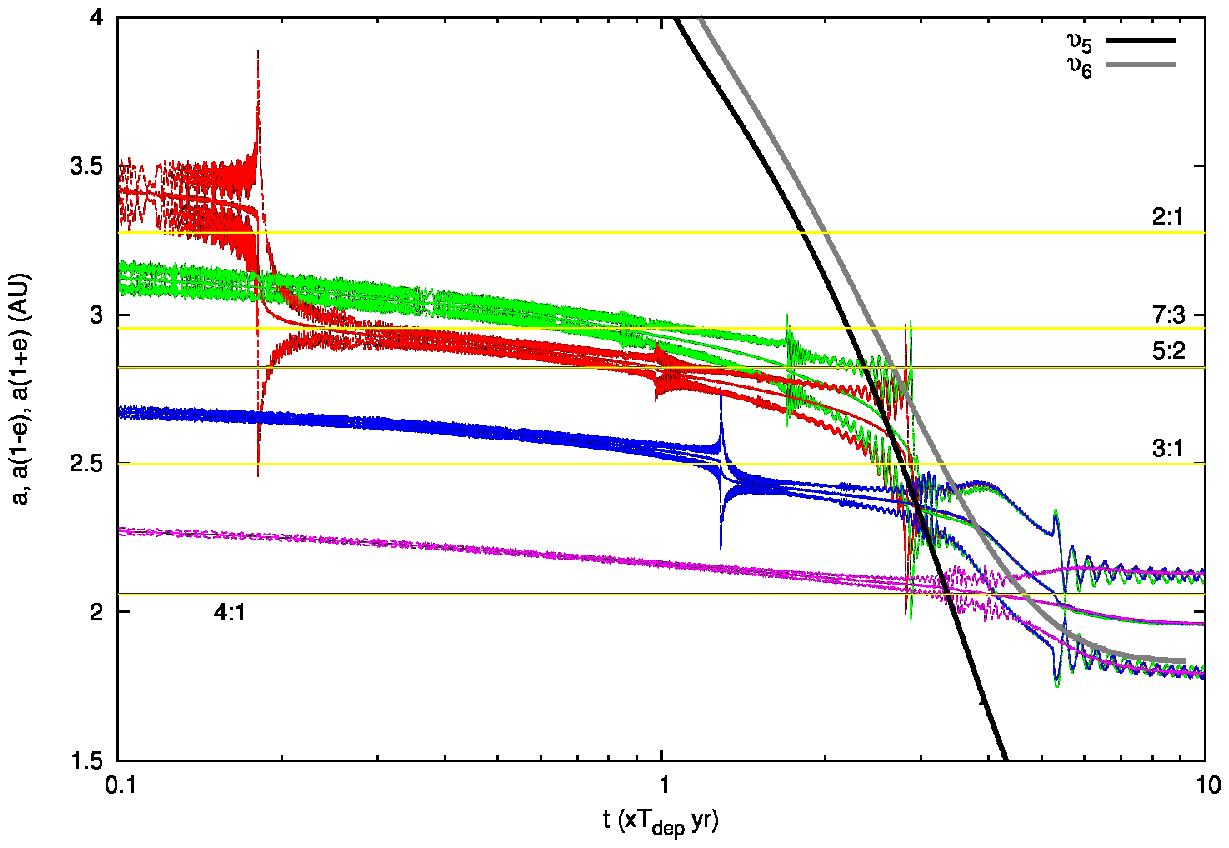}
\caption{The evolution of the perihelion, semimajor axis and aphelion of 
10 km size planetesimals. The black and gray curves represent 
the analytical location of the $\nu_5$ and $\nu_6$ SRs, 
respectively. The yellow lines show the locations of several of Jupiter's 
low-order MMRs.}
\label{fig:a_t_rp10}
\end{figure*}

In order to generalize the results above, we consider other initial $a_p$ values
for two groups of planetesimals, according to their sensitivity to
orbital damping: (i) the strong-damping class and (ii) the weak-damping 
class.  For the strong damping class, we take planetesimals of size 10~km as a representative case. Figure \ref{fig:a_t_rp10} shows 
the evolution of the semimajor axis, perihelion, and aphelion distances,
starting with four different initial semimajor axes,  $a_p = 3.5$~AU, 
$a_p = 3.2$~AU, $a_p = 2.7$~AU, and $a_p = 2.3$~AU.  These locations cover 
the entire region of interest. After time $t=10 T_{\rm dep}$,  all four 
representative planetesimals have undergone orbital decay to regions
closer to the Sun than the main belt region. 

Within $\sim 0.2 T_{\rm dep}$, hydrodynamic drag leads to a fractional 
orbital decay for the planetesimals initially located at 3.5 AU.  
As these planetesimals pass through
Jupiter's and Saturn's MMRs and SRs, their
eccentricities are excited.  Subsequent eccentricity damping leads to 
orbital decay.  Provided that $a_J$ remains constant, the location
of the MMR is fixed.  In the region close to Jupiter ($\ga3$~AU), 
its low-order (especially 2:1) MMRs are primarily 
responsible for the excitation and subsequent inward migration of 
planetesimals \citep{ida1996}.

Closer to the Sun, however, the torque induced by higher-order MMRs is limited by the small magnitude of $e_J$.  The planetesimals'
eccentricity excitation and orbital evolution are dominated by the 
SRs. During disk depletion, the locations of the $\nu_5$ 
and $\nu_6$ SRs are relocated closer to the Sun.
It is therefore possible for some planetesimals to be continually 
excited by the evolving SRs.  After $\sim 5 T_{\rm dep}$,
there is little residual gas left in the disk to induce any significant (i)
precession for Jupiter or Saturn and (ii) eccentricity damping for the 
planetesimals.  Consequently, the propagation of the $\nu_5$ and $\nu_6$ 
SRs is stalled inside the orbit of Venus and outside the
orbit of Mars, respectively.  The planetesimals also retain their eccentricities.

The contributions of the MMRs and SRs can be distinguished 
by the evolution of the planetesimals' semimajor axis and eccentricity.
The planetesimal semimajor axes evolve through several low-order MMRs, including the 
5:2, 3:1, and 4:1 MMRs. In the limit of small $e_J$, the passages
through these other MMRs (apart from the powerful 2:1 MMR) do not lead 
to strong eccentricity excitation and significant planetesimal migration.
Nevertheless, the planetesimals' response to the 3:1 and 4:1 MMRs is enhanced when
the passage is partly coupled with the propagation of the
$\nu_5$ and $\nu_6$ SRs.

\begin{figure*}
\centering
\includegraphics[width=0.95\linewidth,clip=true]{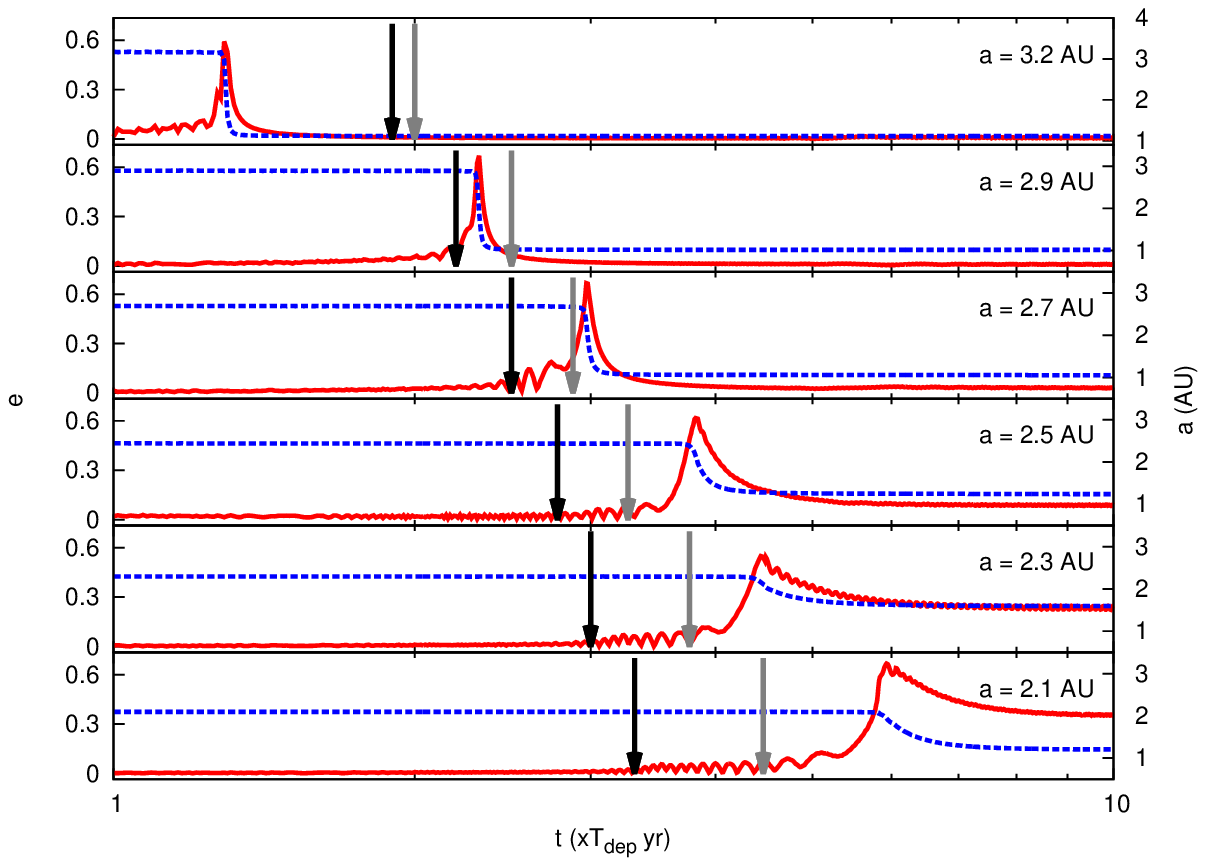}
\caption{Eccentricity (red) and semimajor axis (blue) evolution for 
planetesimals with $r_p = 100$~km. The black/gray arrows indicate the epoch 
of the $\nu_5/\nu_6$ SR passage.}
\label{fig:e_a_t_rp100}
\end{figure*}

For planetesimals in the weak-damping class, the situation is somewhat 
different from the strong-damping class (see Fig.~\ref{fig:a_t_ap3}). 
In Figure~\ref{fig:e_a_t_rp100}, we show 
the orbital evolution of planetesimals in the weak-damping class 
with a representative size, $r_p = 100$~km. These planetesimals' 
orbits evolve under the combined effect of MMRs and the 
sweeping secular resonance. Figure \ref{fig:e_a_t_rp100} shows 
how the $\nu_5$ and $\nu_6$ SRs sweep through the main belt region. 
We compare the analytical solutions for $\nu_5$ (black arrow) and $\nu_6$ 
(grey arrow) SRs with the numerical results 
of \S\ref{sec:nmethod}. Orbital excitation occurs between 2.9~AU and 2.1~AU, 
and eccentricity variations are consistent with the analytical predictions 
for the $\nu_5$ and $\nu_6$ SRs. The only exception is 
the case where the planetesimals are initially located beyond 3~AU ($a_p 
= 3.3$~AU), this region is dominated by the extended 2:1 MMR, and these planetesimals are excited prior to the passage of the SSR.

We can therefore infer that the original location of planetesimals in the main belt region determines whether their excitation is dominated by either the MMR or by the 
sweeping SR, irrespective of the rate of eccentricity damping they experience. In general, the MMR plays a key role in exciting the eccentricity for planetesimals with initial $a_p \ga 3$~AU, 
while for those with initial $a_p \la 3$~AU, the $\nu_5$ and $\nu_6$ SRs are mainly responsible for their angular momentum deficit.


\subsection{Asteroid size selection}

The observed present-day SFD in the main asteroid belt appears to have 
a transition at around $r_p \approx 50$~km \citep[e.g.,][]{bottke2005, 
morbidelli2009}.  This transition has been interpreted in terms of a  
cutoff in the population of small asteroids shortly after their formation.
Such an initial SFD may either be attributed to either the preferential formation
of large asteroids or the clearing of small embryos due to a sweeping 
SR. In this work, we presume that the existence of a 
protoplanetary disk undertakes the natural selection role, rather than 
invoking a new formation mechanism. In this scenario there are two key 
processes at play: (i) planetesimals in the main asteroid belt region 
can be significantly excited as SRs (in this work the $\nu_5$ 
and $\nu_6$ SRs) propagate through the region; and (ii) the excited 
planetesimals experience dispersive inward migration due to the size-dependent 
eccentricity damping force.

\begin{figure*}
\centering
\includegraphics[width=0.95\linewidth,clip=true]{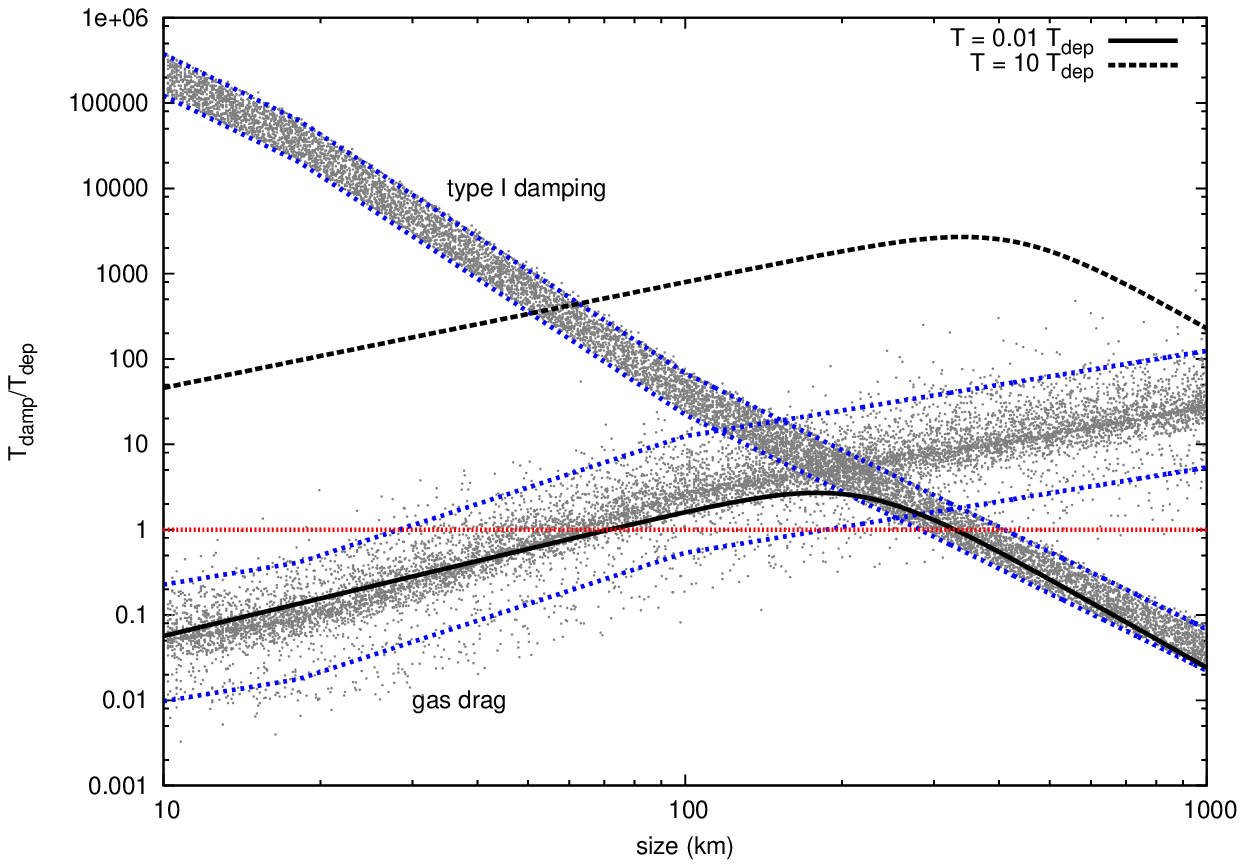}
\caption{Eccentricity damping time (in units of the disk 
depletion time scale) versus planetesimal radius. The two 
mechanisms for eccentricity damping are computed independently. 
When their contribution is combined, the eccentricities of the 
planetesimals are damped on the minimum $T_{\rm damp, tot}$ time 
scale ($T_{\rm damp, tot} = (1/T_{\rm damp, s} + 1/T_{\rm damp,t})^{-1}$). 
The black dots are the simulated results, while the blue dashed curves 
are obtained from analytical fits. The red dashed curve refers to the 
critical value that distinguishes effective and ineffective 
eccentricity damping processes.}
\label{fig:r_t1_t2_td_i_f}
\end{figure*}

In Figure \ref{fig:r_t1_t2_td_i_f}, we show this dependence for an assumed 
initial size distribution $N(r_p) \propto r_p^{-1}$, for a set of 
representative planetesimals with sizes in the range of $10-1000$~km. 
In order to illustrate the size dependence on the two damping mechanisms, 
we analyze the efficiency of these effects separately. The contribution 
from the hydrodynamic drag decreases  as $T_{\rm damp}/T_{\rm dep} 
\propto r_p$ whereas that from the tidal damping follows 
$T_{\rm damp}/T_{\rm dep} \propto r_p^{-3}$ (see eqs.~\ref{eq:t1} 
and~\ref{eq:t2}, respectively). Using these proportionalities, 
we estimate the combined effect of these two damping mechanisms.
The solid black curve shows the damping at $t=0.01T_{\rm dep}$, and 
the dashed black curve indicates the result at the end of the simulation 
($t=10T_{\rm dep}$). Most interesting, the ``bump'' that 
corresponds the weak-damping class shifts to larger-size planetesimals
during the disk depletion.   

\begin{figure*}
\centering
\includegraphics[width=0.95\linewidth,clip=true]{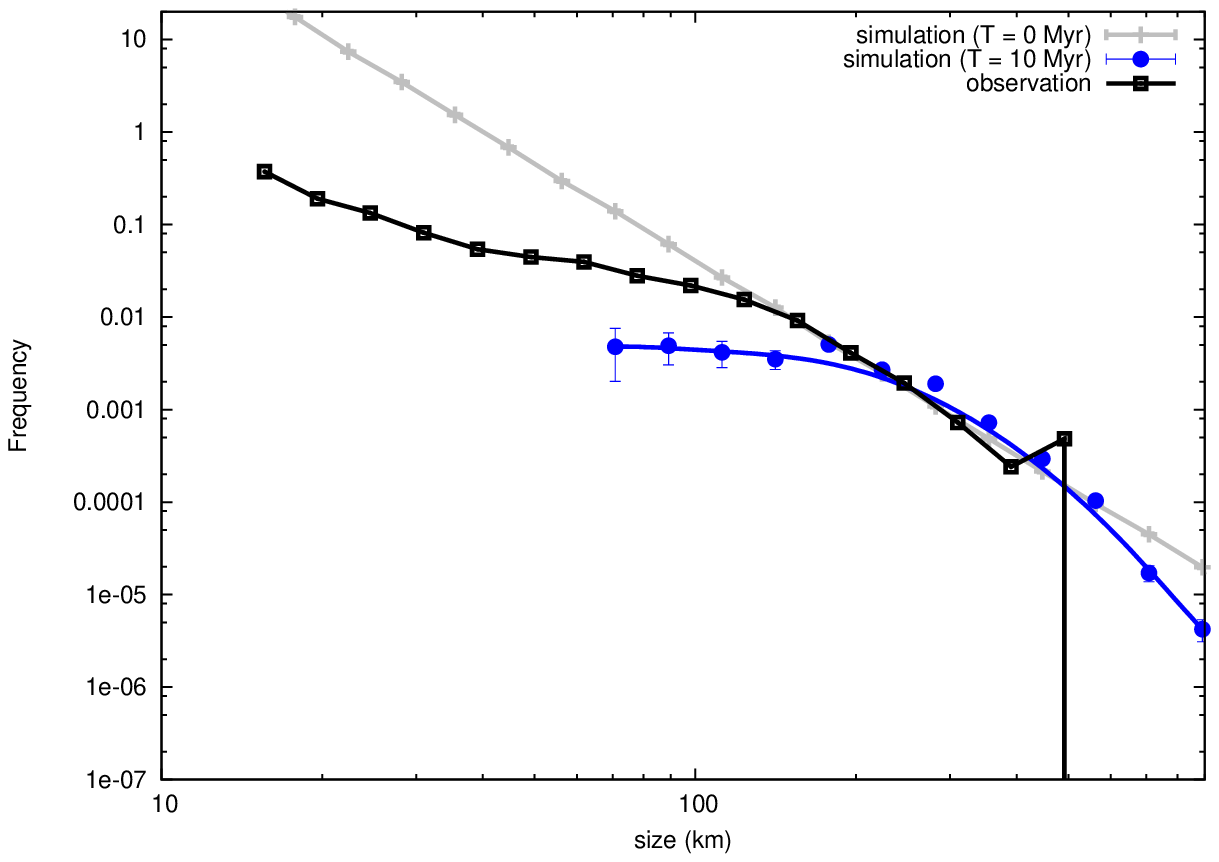}
\caption{Observed SFD and the asymptotic SFD from our 
default model $A_1$ in the main belt region ($\sim 2.1-3.3$ AU). 
The observational data (black curves) are obtained from 
\cite{bottke2005}. The initial ($t = 0$~Myr; gray plus signs) and 
asymptotic ($t = 10$~Myr; blue dots) normalized SFDs are adjusted to fit the observed SFD at $r_p = 200$~km.} 
\label{fig:r_f_obs}
\end{figure*}

A default model $A_1$ is introduced with an initial SFD of the form 
$N(r_p) \propto r_p^{-3.5}$.  While the shape of this SFD 
is retained for the intermediate-size ($300$ km $\la r_p \la$
$1000$ km) planetesimals, the smaller ($r_p \la 50$~km) bodies 
are severely cleared during the phase of gas depletion. The final 
normalized SFD (at $T=10 T_{\rm dep} = 10$ Myr) 
in the main belt region ($\sim 2.1-3.3$ AU) is 
shown in Figure \ref{fig:r_f_obs}. Although the final distribution 
does not completely match the observed SFD, it demonstrates
the possibility of preferential retention of relatively large planetesimals. 
The observed size distribution of the small known asteroids can be produced 
from the subsequent collisional fragmentation process.  There are several
families of asteroids that bear the signature of collisional fragmentation
\citep{zappala2002, nesvorny2006}.

\subsection{Mass depletion in the primordial asteroid belt}

\begin{figure*}
\centering
\includegraphics[width=0.95\linewidth,clip=true]{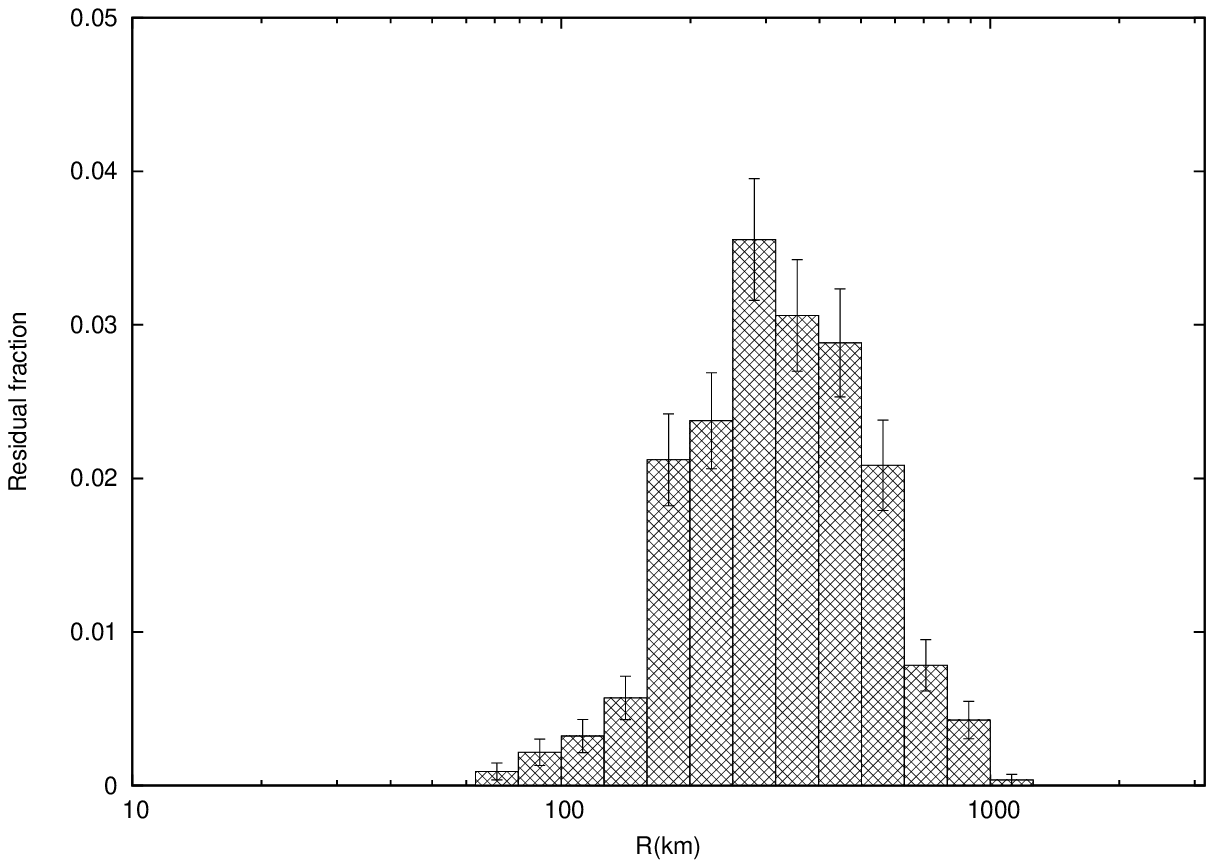}
\caption{Residual fraction of planetesimals in the main belt 
($\sim 2.1-3.3$ AU) as a function of planetesimal size $r_p$ at 
time $t=10 T_{\rm dep}$, for the default model $A_1$.}
\label{fig:r_df}
\end{figure*}

The size dependence of the drag force for eccentricity damping leads to 
differential inward migration rates. 
Moreover, the size range for effective eccentricity damping (i.e., those
planetesimals with $T_{\rm damp}/T_{\rm dep} < 1$) evolves with time. The 
asymptotic retention efficiency (in the main belt)   
indicates that all 
the planetesimals with $r_p \la 50$~km and $r_p \ga 1000$~km are cleared 
and only a small fraction of intermediate-size planetesimals are retained (Fig.~\ref{fig:r_df}). 
These results show that throughout the main belt region, (i) most of the 
planetesimals (99.9 \% in their total mass) are cleared
and (ii) the residual planetesimals have a size 
distribution similar to that of the observed SFD of asteroids larger 
than $r_p \approx 50$~km.  

\begin{figure*}
\centering
\includegraphics[width=1\textwidth]{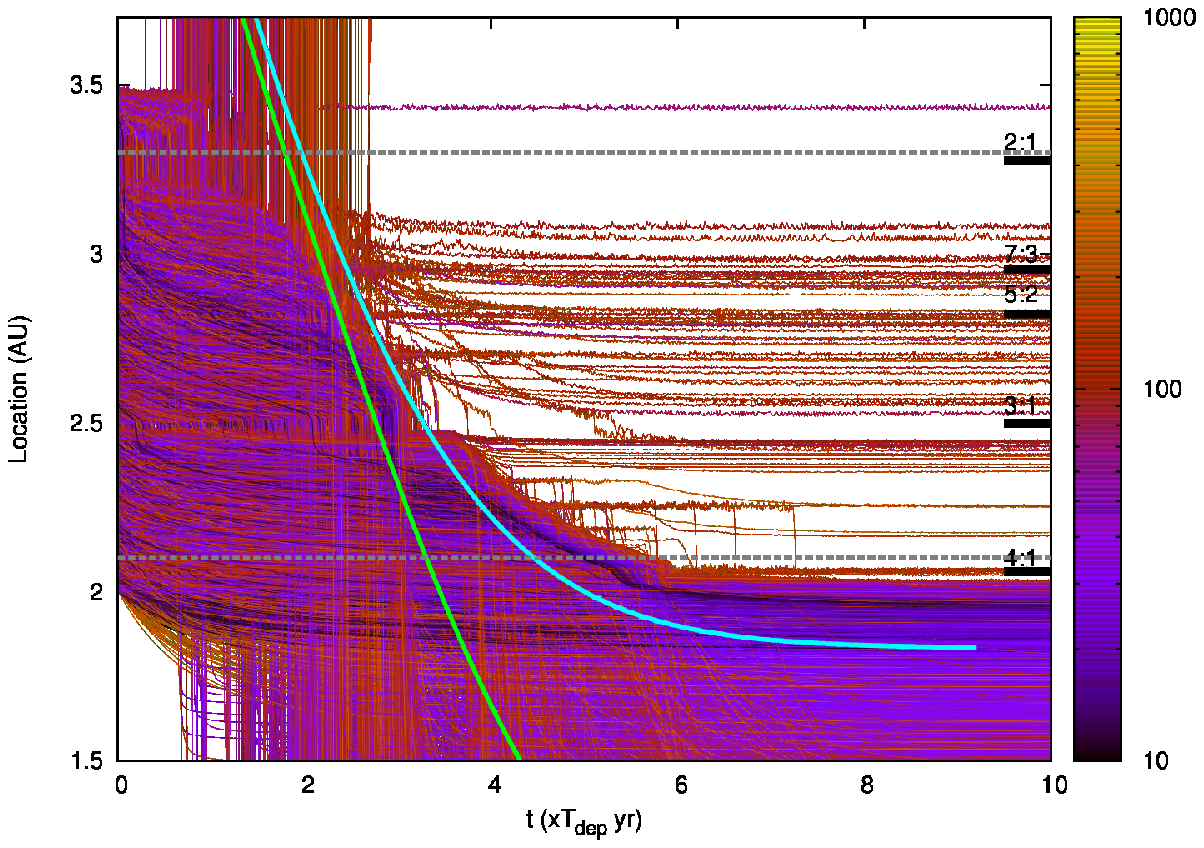}
\caption{Evolution of the semimajor axes within $10\,T_{\rm dep}$. 
The sizes of the asteroids are indicated with different colors, ranging 
from black (10~km) to yellow (1000~km). The green and light-blue curves 
refer to the analytical $\nu_5$ and $\nu_6$, respectively. The gray dashed lines indicate the extent of the present-day main belt. }
\label{fig:a_t_r}
\end{figure*}

The default model $A_1$ also predicts a substantial migration of the
planetesimals with initially $a_p=2-3.5$~AU to the 
region interior to the present-day orbit of Mars (1.5 AU).  
Using the prescription for the evolution of the disk surface 
density distribution in the default model $A_1$, the semimajor-axis 
evolution of planetesimals with $r_p=10-1000$~km
is plotted in Figure \ref{fig:a_t_r}.  A large fraction of this 
population undergoes orbital decay as the $\nu_5$ and $\nu_6$ SRs
propagate inward and excite their eccentricities, which are subsequently damped
by the residual disk gas. As we indicated above, the decrease in
the planetesimals' semimajor axes is due to the combined effects 
of Jupiter's secular perturbation and eccentricity damping by the disk,
provided that the orbital decay time scale $\tau_a = a /{\dot a}
= - \tau_e (1 - e^2)/2 e^2$ is shorter than the $\nu_{5,6}$ SRs'
propagation time scale, $\tau_{\nu_{5,6}} \sim \tau_{\rm dep}$. Therefore, 
after $t \approx 6T_{\rm dep}$, although $\nu_{5,6}$ SRs continue to move 
inward, most of  the planetesimals become detached from the SR recapture.   

We also discuss the asteroids' eccentricity distribution.
In the absence of Jupiter's perturbation and gas drag, 
the residual planetesimals may establish
a collisional equilibrium with a velocity dispersion comparable 
to a fraction of the surface escape speed of the largest members
\citep{aarseth1993, palmer1993, kokuboida1998}.  
If the asteroids' observed eccentricity distribution (up to $e = 0.4$) was
established through their mutual dynamical interaction, it would
require dynamical stirring by a population of large ($> 1000$~km) 
embryos that must be cleared out from the asteroid belt. 

\begin{figure*}
\centering
\includegraphics[width=0.95\linewidth,clip=true]{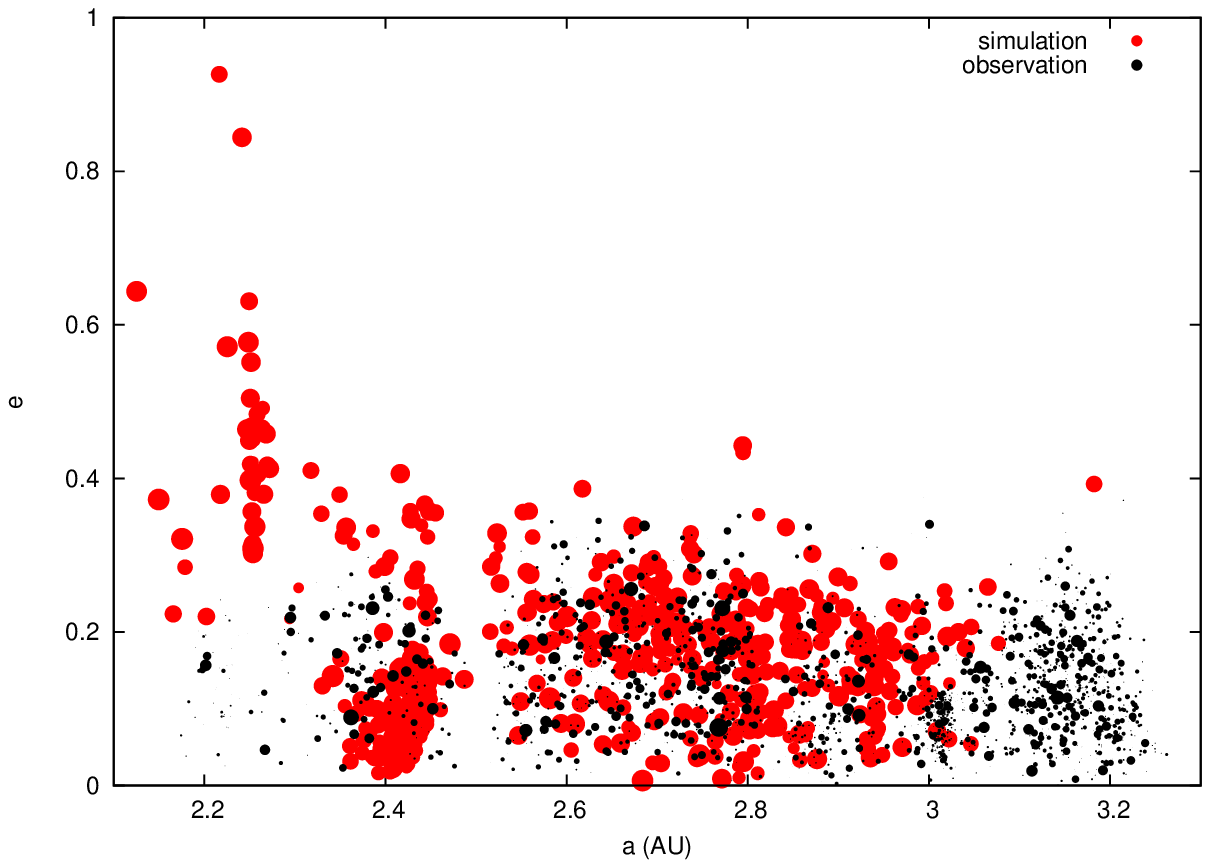}
\caption{Eccentricity of the residual planetesimals (red dots) 
in the main belt ($\sim 2.1-3.3$ AU) as a function of planetesimal 
semimajor axis at time $t = 10\,T_{\rm dep}$, for the default model 
$A_1$.  The semimajor axes and eccentricities of known asteroids are plotted with black dots using the orbit data from the \emph{Minor Planet Center orbit database}.
The dots' size is scaled logarithmically to the planetesimals' radius.  The
labeled dots in the upper right corner correspond to $r_p = 100$ km.}
\label{fig:a_e_r}
\end{figure*}

The SSR can lead to the clearing of 
these large embryos.  Moreover, it can also excite and preserve
modest eccentricity for the retained planetesimals.  In 
Figure \ref{fig:a_e_r} we plot the asymptotic eccentricities of
the surviving planetesimals as a function of their
$r_p$ and $a_p$ in the main belt at time $t=10T_{\rm dep}$. 
As shown above, the surviving planetesimals are 
mainly concentrated in the size range from 100 km to 1000 km, which  
belongs to the weak-damping group. As the eccentricities of most of  
these planetesimals are excited by the SSR, 
their damping timescale is generally longer than the gas disk depletion 
timescale (see also Fig.~\ref{fig:a_t_ap3}).  Most of the planetesimals 
only experience modest orbital decay and maintain an eccentricity 
distribution with a considerable dispersion and an average value 
of $e \sim 0.2$, which is comparable to the observed distribution. 
However, there are several extremely excited cases ($e > 0.6$) 
around $a_p \approx 2.1 - 2.3$~AU.  These planetesimals are able to
retain their high eccentricities  because when the $\nu_{5,6}$ SRs 
sweep through this region after $\sim 5-6$~Myr 
(see also Fig.~\ref{fig:a_t_r}), 
most of the gas material is severely depleted, so that eccentricity 
damping becomes inefficient. These highly eccentric planetesimals 
cross the orbits of Mars, and they are likely to be destabilized 
and ejected from this region.  

The results in Figure \ref{fig:a_e_r} indicate that the eccentricities of the asteroids in the main belt may indeed be due to excitation by the SSRs of Jupiter and 
Saturn.  In contrast, the present-day inclination distribution 
of main belt asteroids cannot be fully attributed to the same mechanism.  Although Jupiter's and 
Saturn's sweeping vertical SR (due to the matching of nodal 
precession rates during the nebula's depletion) can excite the inclination 
of some planetesimals, the magnitude of this effect is too weak to account 
for the observed inclination distribution of asteroids in the main belt \citep{heppenheimer1980, 
ward1981, nagasawa2000}.  Instead, we consider the possibility that  
eccentricity excitation by SSRs induces planetesimals to undergo 
frequent orbit crossings, and that their inclinations were subsequently 
excited to their observed values by their mutual perturbations and 
close encounters.

\begin{figure*}
\centering
\includegraphics[width=0.95\linewidth,clip=true]{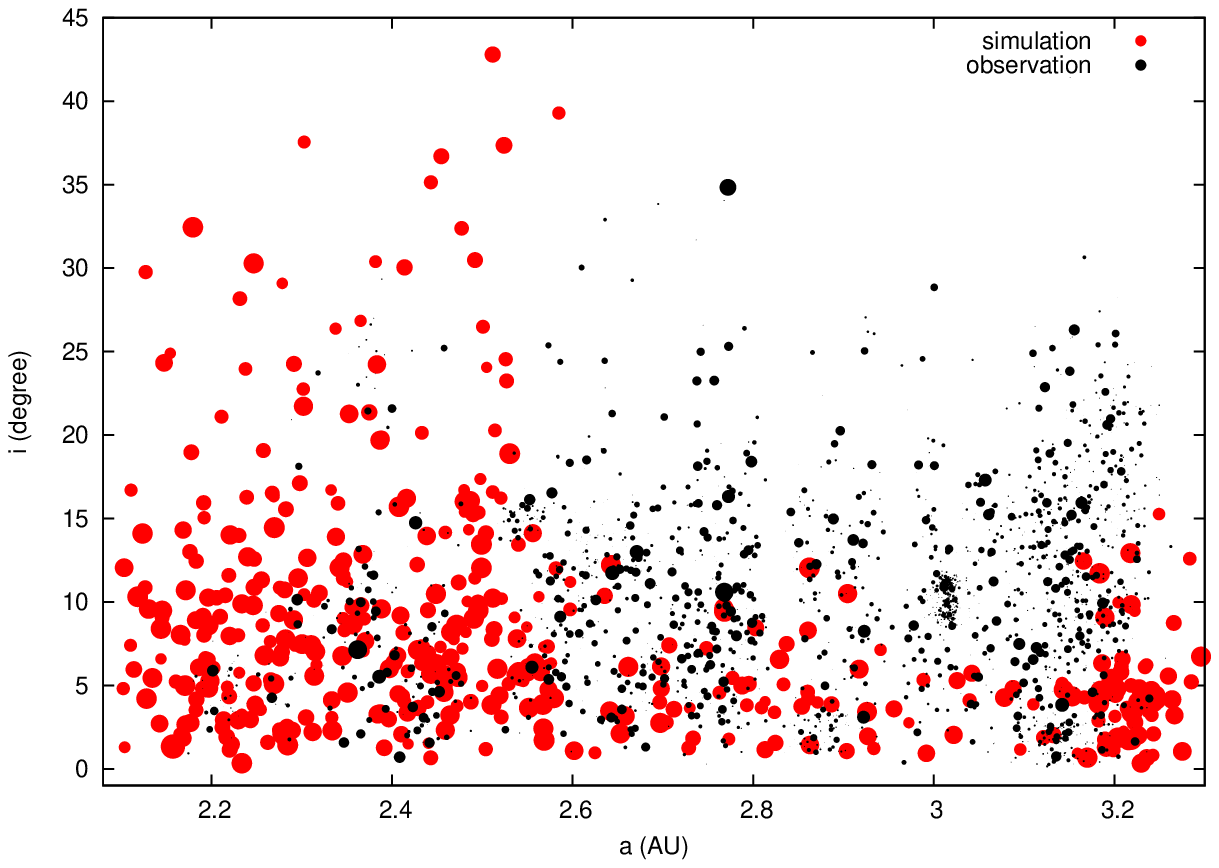}
\caption{Inclination of the residual planetesimals (red dots) 
in the main belt ($\sim 2.1-3.3$ AU) as a function of planetesimal 
semimajor axis at the end of simulation. The semimajor axes and inclinations of known asteroids are plotted with black dots using the orbit data from the \emph{Minor Planet Center orbit database}.
The dot sizes is scaled logarithmically to the planetesimals' radius.  The
labeled dots in the upper right corner correspond to $r_p = 100$ km.}
\label{fig:a_i_r}
\end{figure*}

In order to verify the possibility, we extend our default model with 
100 massless planetesimals plus several large embryos (sizes ranging 
between 1000 and 3000~km), with semimajor axis randomly placed 
between 2.1 and 3.3~AU.  We take into account embryos' gravitational
perturbation on the planetesimals as well as each other.  For 
these intermediate-size planetesimals, inclination
damping due to hydrodynamic and tidal drag is neglected.  
Each of these planetesimals and embryos is assigned with a small 
initial non-zero inclination, randomly chosen from a uniform 
distribution between $i=0^{\circ}$ and $i=5^{\circ}$. We 
carried out 10 sets of simulations.  The inclination distribution 
of our simulated models is compared with the asteroids' present-day 
inclination distribution.  The results in Figure~\ref{fig:a_i_r} 
support our conjecture that although most large embryos are 
cleared out of the main belt region by Jupiter's and Saturn's SSRs, 
the inclination of the retained planetesimals can be 
excited well above their initial values during the migration process, to values 
comparable to that observed among the asteroids.  A sizeable fraction
of planetesimals retains their initial small inclination. This fraction
would be reduced if more embryos were included in
each run so that the frequency of close encounters may be enhanced.
We have neglected the effect of 
nodal precession for Jupiter, Saturn, embryos, and planetesimals.  These 
effects may enhance the inclination excitation for the planetesimals.
Similar to the results in Figure~\ref{fig:a_e_r}, there is a population 
of substantially inclined planetesimals at $a_p \approx 2.1 - 2.3$~AU.  After 
the disk gas has been severely depleted, the orbits of these highly inclined 
planetesimals may be destabilized due to perturbations by Mars.

In Figure \ref{fig:a_t_r} we trace the migration of $10^4$ 
planetesimals and find that the SR starts 
outside of 3.5~AU, which 
is outside the current outer boundary of the main 
asteroids belt, and then sweeps through the entire main belt 
region. Therefore, nearly all the planetesimals in the main belt 
are affected by the $\nu_{5,6}$ SRs as they sweep through
the region. In general, this process results in two possible 
outcomes. First, if their initial semimajor axes are close to that 
of Jupiter, the planetesimals start out in the gap around the planet.
In this limit, Jupiter's and Saturn's secular perturbations are strong and the damping efficiency is due to the relatively 
low surface density of the disk.  Planetesimals in the gap region
are then scattered away from the gas giants. Some planetesimals  
are scattered toward the inner parts of the solar system 
where the eccentricity damping efficiency is high. These planetesimals
then follow the propagation of the $\nu_{5,6}$ SRs. 
A small fraction of the asteroids may be able to avoid being captured
by the sweeping SRs and retain their 
initial location in the main asteroid belt. These are the 
planetesimals that are located relatively far from Jupiter's zone 
of influence and are in the ``preferred'' size range.  

\subsection{Model parameters}

All the results discussed in the previous sections are based on our
default model $A_1$. In this section we briefly discuss how our
choices for the initial conditions and the boundary conditions may influence the final SFDs 
of the asteroids, and additional observations can also help us to put 
further constraints on the properties of the young solar system.

\subsubsection{Dependence on the gas disk}
\label{sec:disk}

In our disk models, we follow the widely adopted MMSN prescription in which
\begin{equation}
H = h_0 \left(\frac{r}{1~\rm AU}\right)^{5/4}  \rm AU \ .
\end{equation}
This result is based on the assumption that gas can establish local 
thermal equilibrium with the solar irradiation everywhere in the disk. 
In the original MMSN model \citep{hayashi1981}, this assumption leads 
to $h_0 = 0.05$.  This assumption is likely to be satisfied after the 
disk becomes optically thin so that most super-micron dust grains and 
planetesimals are directly exposed to the solar radiation.  It also 
requires these dust particles to be thermally coupled to the gas.

However, the inner parts of the solar nebula, including the main belt 
region, may remain opaque to stellar photons even when the surface 
density of the dust grains in these regions is substantially, but 
not severely, depleted from that of the MMSN. In this limit, the gas 
temperature is determined by both viscous dissipation near the opaque 
midplane and stellar irradiation on the exposed dust in the optically 
thin surface layer of a flared nebula.  \cite{garaudlin2007} took 
these effects into account and constructed models with radial 
dependence in the disk thickness and smaller values of $h_0$. 

\begin{figure*}
\centering
\includegraphics[width=1\textwidth]{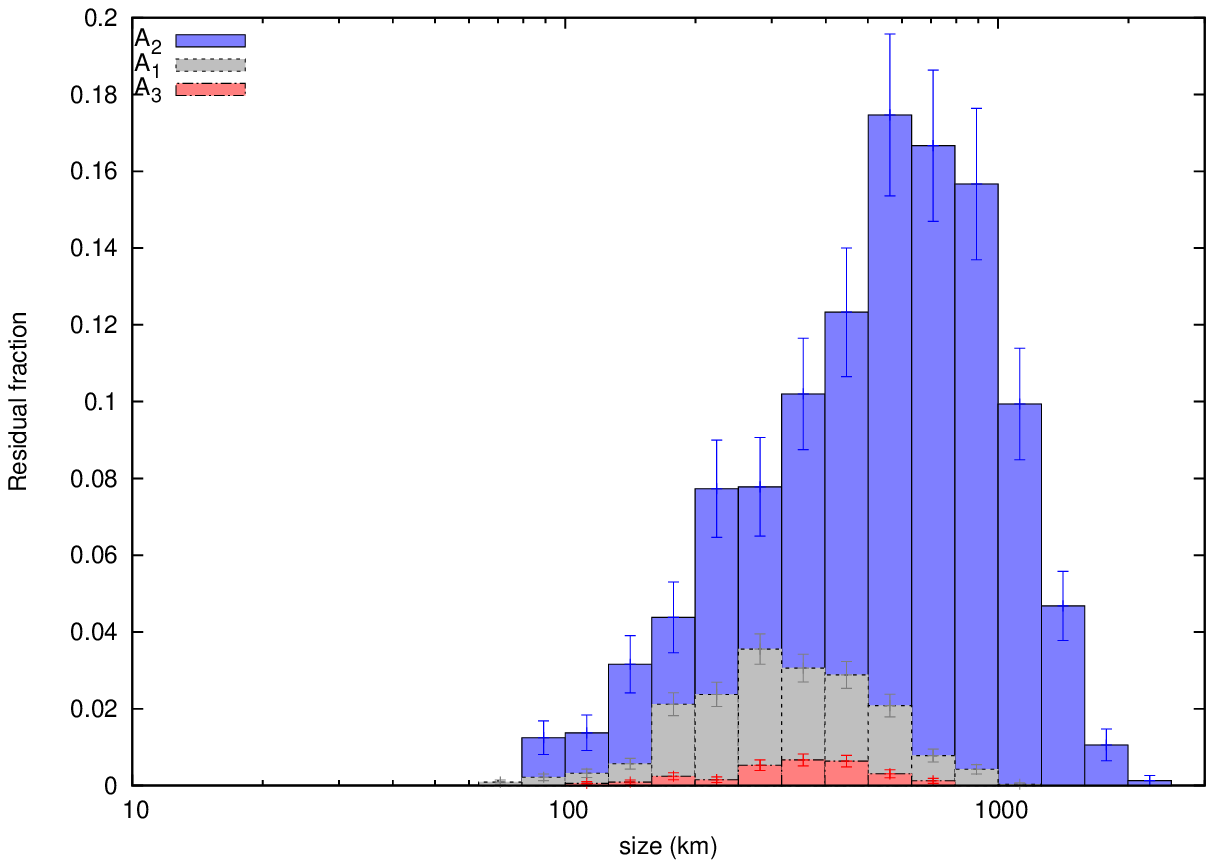}
\caption{Model dependence for disk scale 
heights $h_0$ and the gas depletion timescale $T_{\rm dep}$. The gray histogram shows the results for the default model $A_1$ with $h_0=0.025, T_{\rm dep} = 1$~Myr. The blue histogram shows the results for model $A_2$ with $h_0=0.05, T_{\rm dep} = 1$~Myr. The red histogram represents the results for the model $A_3$ ($h_0=0.025, T_{\rm dep} = 2$~Myr) }
\label{fig:r_df_disk}
\end{figure*}

In our default model $A_1$, we adopt $h_0 = 0.025$, which is 
appropriate for the advanced stages of disk evolution when the main 
belt region remains opaque.  In order to consider the possibility
that the disk may have become optically thin, due to dust coagulation 
and planetesimal formation, we also simulated model $A_2$ with the 
conventional value of $h_0 = 0.05$.  We compare the results of these
two models to evaluate the dependence of $h_0$ on the asymptotic
SFD in the main belt region (Fig.~\ref{fig:r_df_disk}). Since 
$T_{\rm damp,t} \propto H^4$ (eq.~\ref{eq:t2}) and $T_{\rm damp,s} 
\propto H$ (eq.~\ref{eq:t1}), a modification of the scale height factor,
$h_0\propto H$, significantly changes the damping efficiency. 
This dependence is particularly sensitive for  type~I damping, 
and it results in different mass depletion fractions, as well as in a
different asymptotic SFD. As $h_0$ is doubled from model $A_1$ to model
$A_2$, five times more residual planetesimals are retained in the 
the main asteroid belt region. Moreover, the ``bump" in the residual
planetesimals' size distribution also shifts from $\approx 300$~km 
(in model $A_1$) to $\approx 600$~km (model $A_2$). This model dependence
is the result of the tidal damping mechanism for large planetesimals, 
which is most strongly influenced by the value of $h_0$.

The average time scale for the infrared and millimeter excess to decrease
below the detectable threshold is 3-5 Myr \citep{haisch2001}. These features 
are signatures of dust, not gas. Observational data also indicate a wide
dispersion in the disk fading time scale (ranging from 1 to 10~Myr)
even among coeval stars within the same young stellar clusters 
\citep{zuckerman1995}. The gas depletion time scale ($T_{\rm dep}=1$~Myr)
in the default model is on the low end of the observed IR persistent 
time scales, although it is comparable to the shorter time scale inferred 
for the transition from classical to weak-line disks.  Here we consider the
possibility of somewhat longer gas depletion time scales.  In Figure 
\ref{fig:r_df_disk}, we compare the results of the default model $A_1$ to 
those with a gas 
depletion timescale of $T_{\rm dep} = 2$~Myr (model $A_3$). In the latter 
case, fewer asteroids retain their original orbits.  The eccentricity 
damping occurs over a longer period of time, resulting in more extensive
orbital decay from the asteroids' main belt toward the terrestrial 
planet formation region before the gas disk has been depleted. This result 
indicates that if the dispersion time of the gas disk is sufficiently long, 
planetesimal clearing by the SSRs can still be effective even in the limit 
that the gas giant's initial angular momentum deficit is small.

\begin{figure*}
\centering
\includegraphics[width=0.95\linewidth,clip=true]{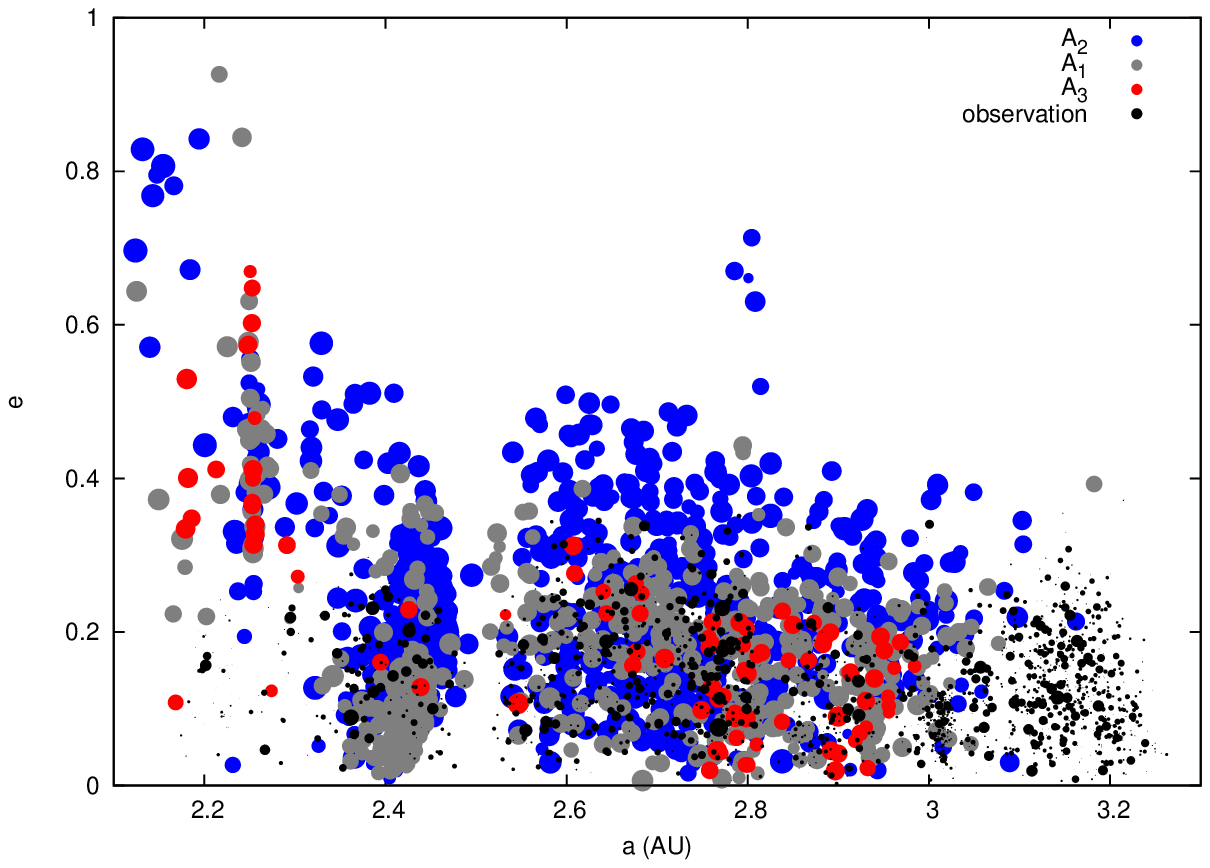}
\caption{Eccentricity vs. semimajor axis and size of the residual planetesimals. The gray dots show the result of default model $A_1$, while the red dots and blue dots are the results for models $A_3$ and $A_2$, respectively. Black dots 
are observed data from the \emph{Minor Planet Center orbit database}. 
The dots' size is scaled logarithmically to the planetesimals' radius.  The
labeled dots on the upper right corner correspond to $r_p = 100$ km.}
\label{fig:a_e_r_disk}
\end{figure*}

The comparison of the residual planetesimals' eccentricity distribution 
between models $A_1$, $A_{2}$, and $A_{3}$ is shown in Figure 
\ref{fig:a_e_r_disk}.  
For model $A_2$, due to the large scale height, more super-1000 km-size 
planetesimals remain in the main belt region, while for model $A_3$, fewer are left. Their eccentricity distribution 
(blue dots and red dots), however, is similar to that of the default model $A_1$, though the
larger planetesimals' (with $r_p \sim 1000$ km) eccentricities are clearly
larger in model $A_2$ than those in model $A_1$, just like the observed 
eccentricity-period distribution of known asteroids (black dots).  
The eccentricities of smaller planetesimals in these models have comparable dispersions and average values.

\subsubsection{Dependence on the gas giants}
\label{sec:gasgiantecc}

\begin{figure*}
\centering
\includegraphics[width=1\textwidth]{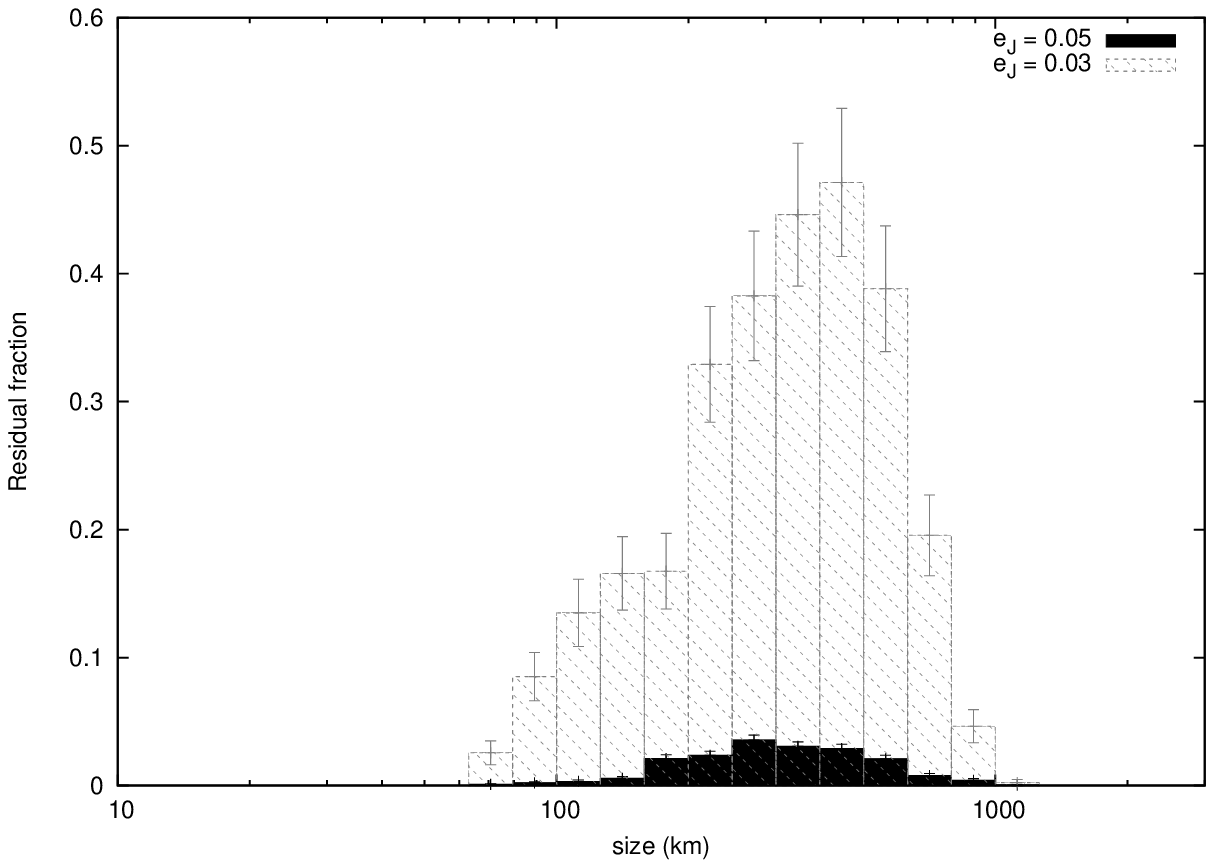}
\caption{The $e_J$ dependence of the retention efficiency and asymptotic 
SFD of asteroids.  The black histogram represents the results for the 
default model $A_1$ ($e_J = 0.05$) and the hatched histogram is for 
$e_J = 0.03$ (model $A_4$).}
\label{fig:r_df_ej}
\end{figure*}

The initial eccentricity of Jupiter 
($e_J$) does not influence the propagation of the sweeping SRs. 
However, $e_J$ does determine the angular momentum deficit of the entire 
system and therefore the strength of the SRs. 
In order to study the effect of Jupiter's eccentricity, we compare 
in Figure \ref{fig:r_df_ej} the results of the default model $A_1$ ($e_J = 0.05$) 
and that of a system with a lower angular momentum deficit, $e_J = 0.03$
(model $A_4$). Note that for the latter model the orbital excitation due to 
Jupiter's SR is weakened, and therefore a larger fraction
of the planetesimals are expected to retain their original orbits. The
less efficient sweeping SRs tend to preferentially 
preserve the smaller bodies rather than the larger Moon-size embryos. 
This result indicates that Jupiter's primordial eccentricity can play 
a substantial role in the clearing efficiency of sweeping SRs. 
In many known exoplanet systems, gas giant planets with
$e_J \gg 0.05$ have been found.  These systems have much larger
angular momentum deficits.  If they attained this angular momentum 
deficit prior to the disk depletion, sweeping SRs 
may have played an important role in clearing the residual 
planetesimals and in inducing the formation of wide dust-free 
gaps in some debris disks 
\citep{su2013}.

\begin{figure*}
\centering
\includegraphics[width=0.95\linewidth,clip=true]{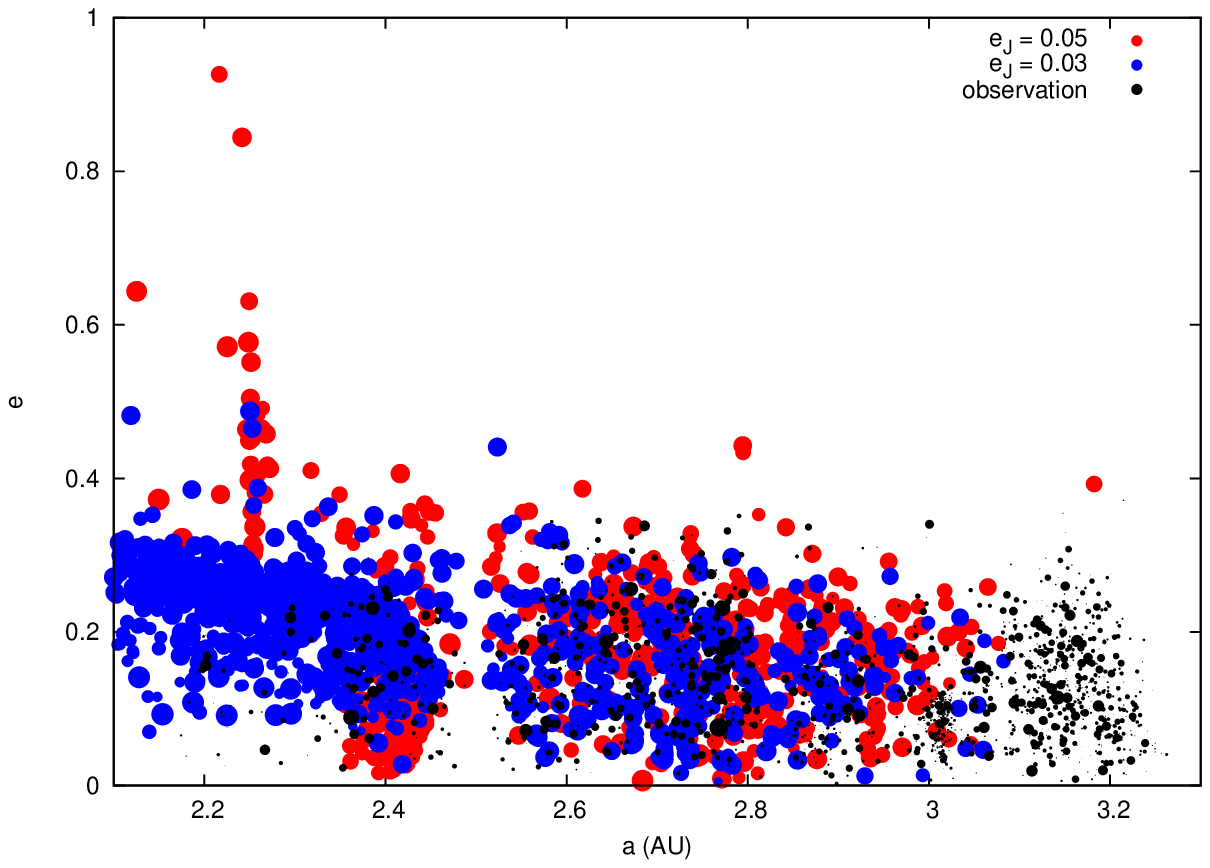}
\caption{Eccentricity as a function of ($a$, $r_p$), for the residual planetesimals, for different eccentricities of Jupiter. The red dots show the results for $e_J = 0.05$, and the blue dots are the results for $e_J = 0.03$. Black dots are observed data from the \emph{Minor Planet Center orbit database}. 
The dots' size is scale logarithmically to the planetesimals' radius.  The 
labeled dots on the upper right corner correspond to $r_p = 100$ km.}
\label{fig:a_e_r_ej}
\end{figure*}

The eccentricity distribution of the residual planetesimals is shown 
in Figure \ref{fig:a_e_r_ej}. We compare models $A_1$ and $A_4$, and 
find that for the smaller $e_J$ it is less likely for the residual 
planetesimals to obtain large eccentricities ($e > 0.6$) due to the 
small angular momentum deficit of the system. However, their average 
values of the orbital eccentricities are similar. 
%

\begin{figure*}
\centering
\includegraphics[width=1\textwidth]{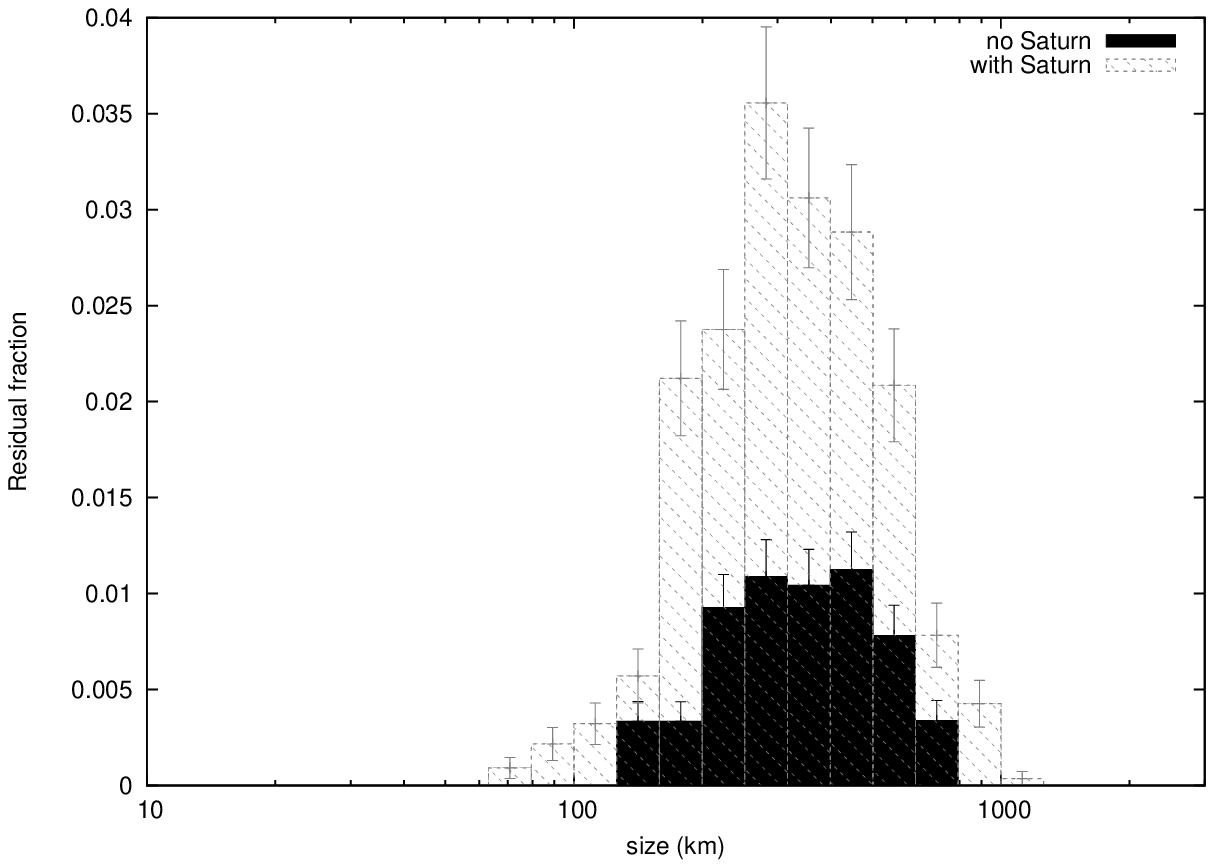}
\caption{Influence of Saturn's secular perturbation on Jupiter. 
The hatched histogram represents the results for the default model $A_1$
(both Jupiter and Saturn included), and the black histogram is for model 
$A_5$, in which the contribution from Saturn is neglected.}
\label{fig:r_df_saturn}
\end{figure*}

In all the models above, we took into consideration perturbation from two gas 
giants, Jupiter and Saturn.  They are the dominant contributors to the 
$\nu_5$ and $\nu_6$ SRs, respectively. However, due to their mass difference, 
Jupiter's SR is generally regarded as the most powerful perturber on 
asteroids' dynamical evolution. In order to isolate the contribution
from the $\nu_5$ SR, we consider an idealized model $A_5$ in which only
Jupiter's perturbation is included. In Figure \ref{fig:r_df_saturn}, the
results of model $A_5$ are compared with those generated with the default 
model ($A_1$). These results show that the $\nu_5$ SSR alone is more effective
in clearing the main belt region and inducing the size selection for the 
residual planetesimals. In the absence of Saturn, the angular momentum deficit of
Jupiter's orbit is preserved. Consequently, $e_J$ preserves its initial value
rather than modulating on a secular time scale (due to the perturbation on 
Jupiter by Saturn).  Therefore, a larger fraction of the initial 
population of asteroids is cleared out in model $A_5$ than in model $A_1$.

We now consider the possibility that the orbits of Jupiter and Saturn may
have evolved after the phase of gas depletion. This possibility has been suggested
by \citet{gomes2005} in the Nice model for the late heavy bombardment (LHB).  This 
scenario is based on the assumption that Jupiter and Saturn were closer 
to each other and their orbits evolved into their present-day configuration
as they scattered and cleared away residual planetesimals in the outer
solar system \citep{fernandez1984}. This hypothesis provides a natural 
explanation for the origin of MMR between Neptune and 
some Kuiper belt objects, including Pluto \citep{malhotra1993}.  However,
there are alternative models based on the assumption that Pluto and several
Kuiper Belt objects were captured into Neptune's MMRs
as a consequence of Neptune's outward migration induced by its tidal 
interaction with the residual gas in the outer regions of the solar nebula
\citep{ida2000}. Similar mechanisms have been proposed to account for several
resonant extrasolar planetary systems \citep{lee2002}.
Such a scenario would not require extensive post-formation 
migration for Saturn.  In view of these possibilities, we consider 
several variations of our default model.  In models $A_6$, $A_7$, $A_8$, 
and $A_9$, we assume Saturn's semimajor axis, at the epoch of nebula 
depletion, to be 8.6~AU, 9.0~AU, 9.3~AU, and 8.1~AU, respectively.  
Without the loss of generality, we set Jupiter's semimajor axis to be 5.2~AU.

The secular interaction between Jupiter and Saturn modifies their precession 
frequencies due to the disk potential and detunes their contribution to the
$\nu_5$ and $\nu_6$ SRs on the residual planetesimals.  The intensity of this 
interaction would be enhanced if they would have had a smaller $a_S - a_J$ in
the past.  In such a limit, the $\nu_5$ and $\nu_6$ SRs would be less 
effective in exciting the planetesimals' eccentricities.  A closer initial
separation between Jupiter and Saturn also would reduce SRs' 
sweeping speed and causes the SRs to be stalled at larger $a$ value during
the epoch of nebula depletion.

\begin{figure*}
\centering
\includegraphics[width=1\textwidth]{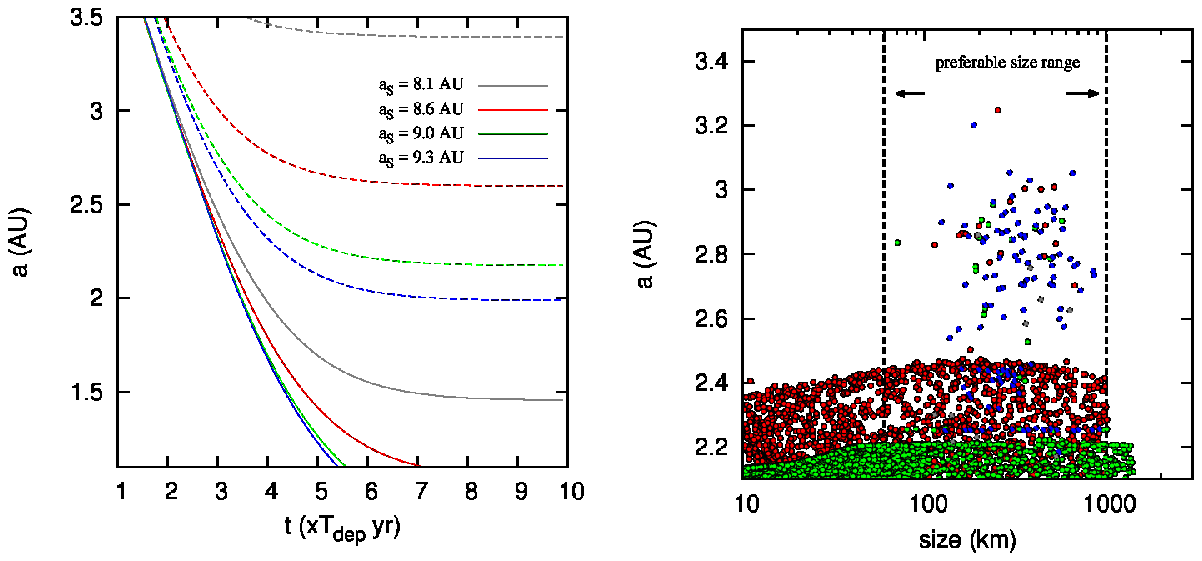} 
\caption{Comparison between the results for models with different Jupiter-Saturn separations. 
The left panel shows the sweeping paths of $\nu_{5}$ (solid 
curves) and $\nu_6$ (dashed curves). We adopt the present-day 
location of Jupiter. Colors label the cases in which the initial 
semimajor axis of Saturn is 8.6~AU (red: model $A_6$),  9.0~AU 
(green: model $A_7$),  9.3~AU (blue: model $A_8$) and 8.1~AU 
(gray: model $A_9$). The right panel shows the planetesimals' 
locations at $t = 10$~Myr as a function of size, and as a function 
of Saturn's initial location (models $A_{6}- A_{9}$).}
\label{fig:v56_r_af}
\end{figure*}

\begin{figure*}
\centering
\includegraphics[width=1\textwidth]{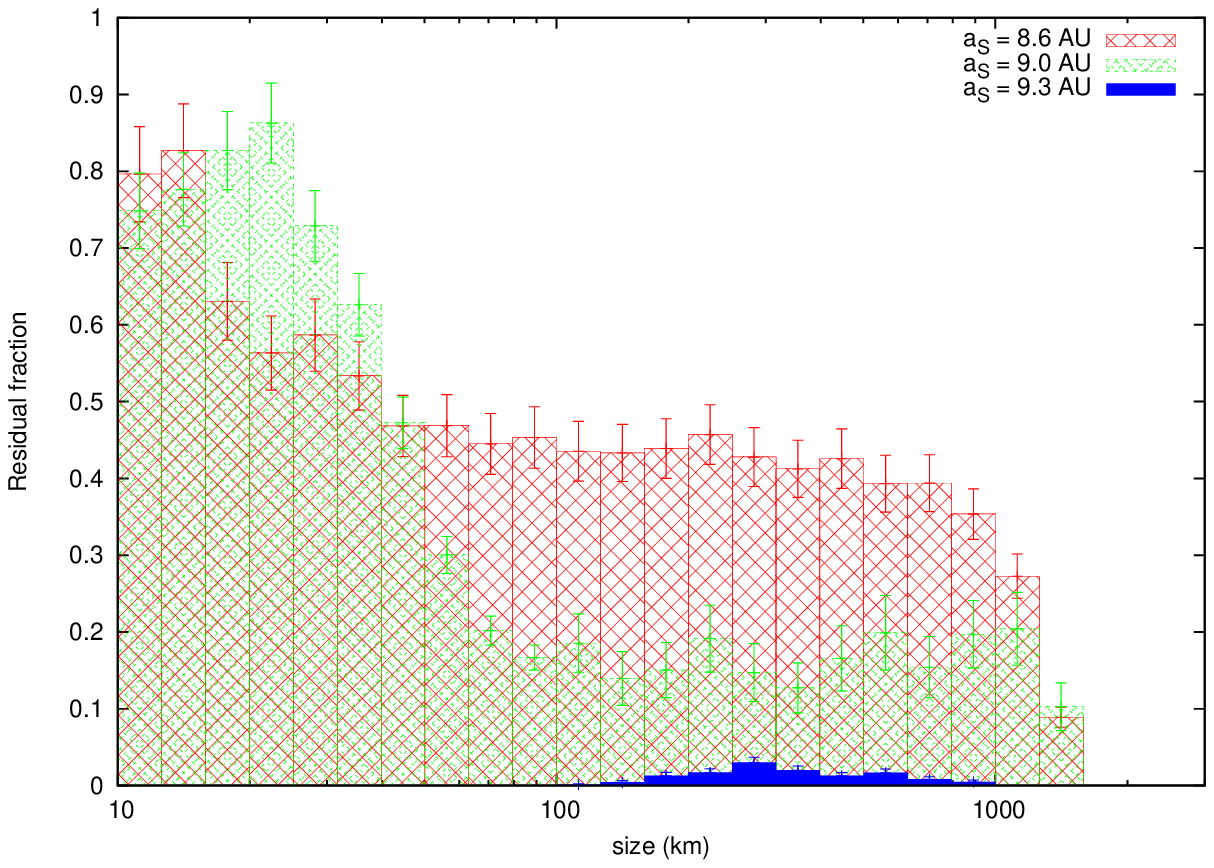}
\caption{Fraction of an initial population of different-size 
planetesimals retained in the main belt ($2.1-3.3$~AU) region in 
models $A_6$ (red), $A_7$ (green), and $A_8$ (blue). These results indicate
that a much larger fraction of residual planetesimals of
all sizes may be retained near the inner boundary of the 
main belt due to the stalling of the $\nu_6$ SR if Saturn 
was located at $a_s = 8.6$~AU or $9.0$~AU at the epoch of 
nebula depletion.}
\label{fig:r_df_ns}
\end{figure*}

Since $M_J \gg M_S$, modification of the $\nu_5$ SSR in models $A_6 -
A_9$ from model $A_1$ is relatively limited such that it is still effective
in clearing out a large fraction of the initial planetesimal population.  
However, the impact of the $\nu_6$ SSR depends more sensitively on $a_S - a_J$
(Fig.~\ref{fig:v56_r_af}).
For example, in model $A_6$ (represented by red dots), the $\nu_6$ SR
is stalled at $\sim 2.6$ AU, and a significant fraction ($\sim 40\%$ in 
total mass) of the initial planetesimal population 
(Fig.~\ref{fig:r_df_ns}), over a wide size range, is retained interior 
to $\sim 2.5$ AU. Some of these planetesimals were initially located
in the region between the $\nu_5$ and $\nu_6$ SRs at the onset of disk
depletion.  Due to the stalling of the $\nu_6$ SR, a similar population 
 ($\sim 20\%$ in total mass) of residual planetesimals is 
retained inside $\sim 2.2$ AU in model $A_7$ (represented by green dots).  
In model $A_8$, the $\nu_6$ SR is stalled inside 2 AU and most planetesimals 
are cleared out of the main belt region. Only a small fraction (comparable
to that of the default model $A_1$) of the 
initial planetesimals within a selected size range  ($50-1000$ km) is 
retained (see 'preferable size range' in Figure ~\ref{fig:v56_r_af}) 
in the classical main belt region ($\sim 2.1-3.3$~AU).  

\begin{figure*}
\centering
\includegraphics[width=1\textwidth]{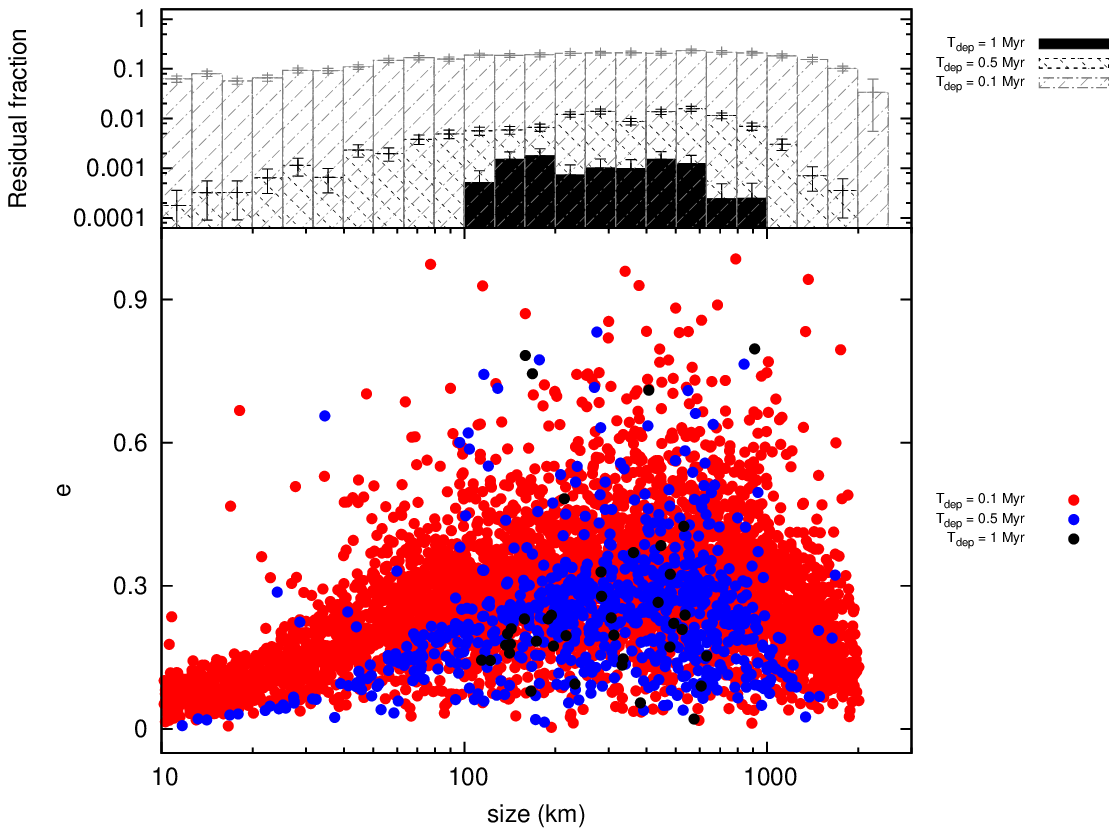}
\caption{Top: Fraction of an initial population of different-size 
planetesimals retained in the main belt ($2.1-3.3$~AU) region in 
models $A_9$ with $T_{\rm dep} = 1$Myr (solid black), 0.5 Myr
(short dashed), and 0.1 Myr (long dashed). Both the fraction 
and size range of the residual planetesimals decrease with 
$T_{\rm dep}$.  Bottom: asymptotic eccentricity of different-size 
residual planetesimals in models $A_9$ with $T_{\rm dep} = 1$Myr 
(black), 0.5 Myr (blue), and 0.1 Myr (red).  The correlation
between the planetesimals' retention fraction and $T_{\rm dep}$ 
is determined by the efficiency of eccentricity damping. For 
relatively short $T_{\rm dep}$ ($<0.5$ Myr), a fraction of 
intermediate-size planetesimals have sufficiently eccentricity 
for them to cross the orbits of terrestrial planets and become 
dynamically unstable. The asymptotic eccentricity of all residual 
planetesimals decreases with $T_{\rm dep}$. }
\label{fig:r_df_tdep}
\end{figure*}

When the semimajor axis of Saturn is reduced from 8.6~AU (model 
$A_6$) to 8.1~AU (model $A_9$), the influence of the 2:1 mean motion resonance 
between Jupiter and Saturn can no longer be neglected. The secular precessions 
of both Jupiter and Saturn are significantly altered, especially when they 
are near the 2:1 MMR location at $\sim 8.25~$AU  
\citep{malhotra1989, murray1999, agnorlin2012}.  In model $A_9$, the 
propagation of the $\nu_6$ SR is stalled outside $\sim 3.35$~AU, while 
the $\nu_5$ SR still sweeps through the main belt region, albeit at a 
slower pace. The dynamical evolution of the planetesimals is similar
to that in model $A_5$ (with Jupiter only).  The stagnation of the $\nu_6$
SR prevents the accumulation of a significant population of residual
planetesimals in the main belt region.  The fraction of the initial 
population of planetesimals ({$~10^{-4}$ in mass}) retained in   
model $A_9$ is an order of magnitude smaller than that of the default 
model $A_1$.  The total mass of the residual planetesimals in this 
case is also smaller than that of the present-day asteroid belt.  The residual planetesimals would not be able to provide
a reservoir of asteroids to accommodate the LHB
induced by Saturn's subsequent passage through Jupiter's 2:1 MMR (in accordance with the Nice model) unless many more 
residual planetesimals may be retained with a much shorter ($ \ll 1$~Myr) 
nebula depletion time scale {\bf (Fig. \ref{fig:r_df_tdep})} .
These results indicate that during the epoch of disk depletion, 
efficient clearing of residual intermediate-size planetesimals requires
either the separation of Jupiter and Saturn ($a_s-a_J$) to be 
within $\leq 0.5$ AU of its present-day value (i.e., the migration of
both gas giants after the gas depletion was limited) or that they were 
located near their 2:1 MMR. For the latter 
possibility, the clearing of residual planetesimals may be overly
efficient.

\begin{figure*}
\centering
\includegraphics[width=1\textwidth]{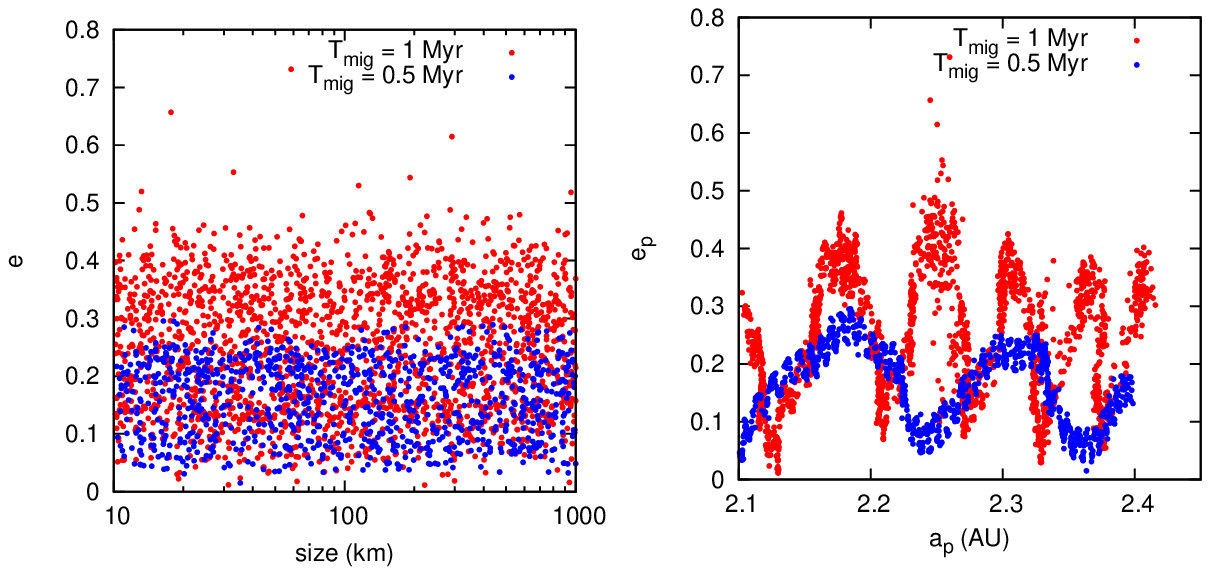}
\caption{Kinematic properties of the residual planetesimals resulting from
Saturn's migration after gas in the nebula is depleted. The 
semimajor axis and eccentricity distributions of the planetesimals are obtained from model $A_6$. 
Saturn is assumed to have migrated outward from $8.6$~AU to its present-day
orbit over a time scale $T_{\rm mig}=0.5$~Myr (blue dots) and 1~Myr (red dots).  
Left panel: the asymptotic eccentricity of residual planetesimals
in the main belt region. Right panel: The semimajor-axis-eccentricity 
distribution of the residual planetesimals. The planetesimals' semimajor 
axes do not change significantly during Saturn's outward migration,
but their orbits can become dynamically unstable due to the perturbations
by both the terrestrial planets and the ice giants.}
\label{fig:r_df_e}
\end{figure*}

In models $A_6$ and $A_7$, Saturn still has to undergo further orbital 
migration to its present-day kinematic configuration after the gas 
depletion phase. During this late phase of orbital evolution, the eccentricities
of the residual planetesimals are excited in the absence of both 
hydrodynamic and tidal damping. Their semimajor axes are not affected
by the SSRs.  Nevertheless, their enhanced eccentricity may lead to 
orbit crossing with the terrestrial planets and subsequent dynamical instabilities.
In order to consider this possibility, we simulated the dynamical
evolution of the residual planetesimals (mainly located between 2.1~AU 
and 2.4~AU, size of 10 to 1000~km) in model $A_6$ with an outward
migration of Saturn over two different time scales (1 and 0.5~Myr).  
In the left and right panels of Figure~\ref{fig:r_df_e}, we plot the 
asymptotic eccentricity for different-size planetesimals in the main 
belt region. During Saturn's migration, $\nu_6$ sweeps from $\sim 2.6$
to  $< 2.1$~AU.  Planetesimals retained in this region after the nebula
depletion are further excited. In the absence of any effective
eccentricity damping, most of the residual planetesimals retain 
somewhat larger eccentricities. And the maximum amplitude of those 
eccentricities is correlated with the migration time scale of Saturn. 
Protracted migration (on a time scale $>1$ Myr) generally excites
the eccentricity of the residual planetesimals above that of the observed asteroids ($e < 0.3$). The eccentricity of some residual 
planetesimals may be sufficiently large for them to cross the orbits 
of the terrestrial planets and become dynamically unstable.  In a 
gas-free environment, size selection among the retained planetesimals 
is no longer possible.

In models $A_{6}-A_9$, we have not considered the possibility 
that Jupiter may have migrated as extensively as hypothesized in
the ``grand-tack'' model \citep{walsh2011}.  In such a scenario, most of 
the planetesimals formed in the main belt region would be scattered into 
the terrestrial planet zone and the $\nu_5$ and $\nu_6$ SSRs would have little 
impact on the essentially empty main belt region during the depletion of
the disk gas. The migration paths of multiple planets depend sensitively 
on the initial conditions and boundary conditions of the solar nebula, as well as the 
time lapse between Jupiter's formation and Saturn's formation and their gas accretion 
rates.  In light of these uncertainties, we consider here an alternative 
possibility under the assumption that Jupiter and Saturn's migration 
was limited in range. Our results confirm that $\nu_5$ and $\nu_6$ SSRs 
alone can lead to the extensive inward migration of the planetesimals 
from the main belt region to the terrestrial planet zone. They also show that
the intermediate-size planetesimals are preferentially retained, regardless 
of the initial value of $a_S$.


\section{Summary and discussion}

We have constructed a sweeping SR model to explain the observed
properties of the main asteroid belt without the necessity of introducing
additional assumptions about the origin and evolution of planetesimals. Using 
this model, we are able to reproduce (i) the apparent mass deficit of 
the present-day main asteroid belt \citep[Fig.~\ref{fig:r_df}; see also][]{morbidelli2009}, (ii) the observed SFD 
of the asteroids in the main belt (Fig.~\ref{fig:r_f_obs}), (iii) the 
semimajor axis and eccentricity distributions of residual planetesimals 
(Fig.~\ref{fig:a_e_r}), and (iv) substantial radial mixing 
(Fig.~\ref{fig:a_t_r}).

Some of these issues have already been considered previously by several 
other investigators, especially those associated with the Nice school. 
In this summary and discussion section, we make several detailed comparisons 
between our results and those obtained with other promising models. We 
highlight here some tension between various models and suggest that 
our sweeping SR hypothesis offers a viable alternative, at least in 
the context of asteroids' size-frequency and spacial distributions.

\subsection{Comparison with the ``{\it ab inito} large" model for the asteroids' SFD}
\label{sec:sum1}

We have seen in this paper that planetesimal eccentricities can 
be excited through the exchange of angular momentum with Jupiter and 
Saturn through their MMRs and SRs, and that eccentricity damping leads to 
energy dissipation and the orbital decay of planetesimals. In our models,
we assume that the surface density of the nebula decreases everywhere
with a radius-independent depletion factor.  We show that even after 
the disk gas is sufficiently depleted to enable the $\nu_{5,6}$ SRs
to sweep through the main belt region, eccentricity damping continues
to be effective in causing some planetesimals to undergo orbital decay.
This eccentricity damping process is particularly effective for the 
small planetesimals, due to the hydrodynamic drag, and for the 
large embryos, due to their tidal interaction with gas in the solar 
nebula. The competition between these two damping mechanisms leads 
to planetesimals' size-dependent orbital decay rates. The similarity 
between the size distributions of the residual planetesimals and 
the observed SFD provides an alternative scenario for the origin of 
the asteroids. In contrast, \citet{morbidelli2009} propose an ``{\it ab initio} large" model that requires the minimum size of the primary planetesimals, $\sim 100$~km in diameter, to reproduce the observed ``bump" feature in the present-day SFD of the asteroids; however, a new planetesimal formation scenario is required in this model.

\subsection{Comparison with previous models for the planetesimal
removal from the main belt region}
\label{sec:sum2}

In a related analysis, \cite{obrien2007} showed 
that the SR sweeping can neither excite the eccentricity nor 
efficiently clear the residual planetesimals. The main difference
between their and our results is mainly due to their neglect of  
any damping of the planetesimals' excited eccentricities. 
They justified this approximation with a scenario in which the solar 
nebula may have been cleared outward from its center.  They also 
considered the SR sweeping effect with a radius-independent 
depletion model, in which there would be a non-negligible amount 
of residual gas in the main belt region.  However, they did not 
explicitly compute the contribution of this residual gas on the 
eccentricity damping for the most vulnerable planetesimals
in the main belt region. 

For the orbital parameters of Jupiter and Saturn, \cite{obrien2007} 
adopted (1) a Nice model with negligible eccentricity \citep{gomes2005} 
and (2) their present-day values. With nearly circular orbits, the 
amount of angular momentum Jupiter and Saturn can receive would be 
limited.  They found that the excitation of planetesimals' 
eccentricities to the present values of the asteroids would require 
a slow propagation of the SRs with an implied gas-depletion time 
scale $>10$ Myr.  With the present-day eccentricities of Jupiter 
and Saturn, planetesimal eccentricities can be adequately excited 
within a few megayears.  However, since the SRs do not directly lead 
to energy transfer, they found that most planetesimals' semimajor axes would not be affected by the passage of Jupiter's and 
Saturn's SRs.  

Based on the discussions in \S\ref{sec:quantitive}, we adopted in our
default model $A_1$, an angular momentum deficit for the Jupiter-Saturn configuration 
to be 2/3 of its present-day value.  Similar to the \cite{obrien2007} 
present-day model, our model $A_1$ reproduces the observed eccentricity-semimajor-axis distribution of the residual intermediate-size planetesimals. 
When a quasi-equilibrium in eccentricity excitation/damping is established, 
Jupiter's and Saturn's MMRs and sweeping SRs relocate $>99\%$ of the 
initial planetesimals in the main zone region to regions interior to 2 AU. 
These contrasting conclusions on whether the sweeping SRs can 
remove most of the residual planetesimals from the main belt region are determined by
the assumed efficiency of eccentricity damping.  

\subsection{Comparison with the Embedded-embryos model for 
the planetesimal eccentricity excitation}

In the core accretion scenario, the formation of gas giant 
planets is proceeded by the emergence of protoplanetary 
embryos with masses comparable to or larger than that of the
Earth \citep{pollack1996,idalin2004}. These embryos are 
embedded among residual planetesimals in the gas-rich nebula.
\cite{wetherill1992} suggested that the eccentricity and 
inclination of embryos and planetesimals in the main belt 
region may be excited by Jupiter's MMRs and SRs.  This 
hypothesis was confirmed by a series of numerical simulations 
\citep{chambers2001, petit2001} which showed that through 
encounters between them, planetesimals may be cleared from 
the main belt region,
while the more massive embryos are retained in the main
belt region.  

In the embedded-embryo scenario, 
\cite{obrien2007} showed that up to $\sim 90-95\%$ of the 
residual planetesimals may be cleared over at least several megayears. 
Despite this remarkable reduction, the surface density of the 
residual planetesimals would still be an order of magnitude 
larger than that of the present-day asteroid population.  
One possible scenario to make up this difference has been 
suggested on the basis of the Nice model 
for LHB.  During the early stages 
of solar system evolution, it is likely that Saturn may 
have migrated outward as it scattered planetesimals in 
the outer solar system to even larger heliocentric 
distances \citep{fernandez1984, malhotra1993}. The Nice 
LHB model is based on the assumption that Saturn has migrated 
over $\sim 2$ AU. It attributes the LHB event and the 
capture of Trojan asteroids to the 
passage of Saturn's orbit through its 2:1 MMR with Jupiter 
at $\sim 3.8$ Gyr ago. At that time, the surface density
of the residual planetesimals in the main belt region may be 
reduced by an order of magnitude due to the clearing of the 
sweeping SRs \citep{gomes2005, morbidelli2005, tsiganis2005}.

Our model adopts some similar assumptions to the embedded-embryo 
model, such as 
an initial planetesimal size distribution that includes the existence
of relatively massive embryos. We also adopt a subset of the initial
conditions assumed by the Nice model.  For example, we only consider 
a radius-independent depletion factor (\S\ref{sec:sum1}), and we adopt
a modest angular momentum deficit rather than nearly circular orbits
for the Jupiter-Saturn system (\S\ref{sec:sum2}). With our default model
$A_1$, we showed that as they sweep through the solar nebula, the $\nu_5$ 
and $\nu_6$ SSRs can excite planetesimals' eccentricities. Effective 
hydrodynamic and tidal drag damps planetesimals' eccentricities and induces
them to undergo orbital decay.  Our results suggest that most of the
small asteroids and large embryos are removed from the main belt region
during the first few megayears of solar system evolution.  

Based on previous investigations\citep{heppenheimer1980, ward1981, 
nagasawa2000}, we recognize that asteroids' present-day inclination
cannot be excited by Jupiter's and Saturn's secular perturbation 
during the nebula's depletion because their inclination resonances 
do not sweep through the main belt region.  We showed with the parameters
of model $A_1$, that the mutual perturbations between planetesimals 
(with sizes in the range of $100-1000$~km) and a few embryos 
(with sizes $1000-3000$~km) can adequately excite the inclinations 
of intermediate-size planetesimals to observed values (Fig.~\ref{fig:a_i_r}).
We suggest that these planetesimals are most likely to be the progenitors of 
the asteroids as they are preferentially retained in our model.

\subsection{Comparison with the Grand-tack model for 
the removal of embryos from the main belt region}

An implication of the embedded-embryo scenario is that 
a population of relatively massive embryos needs to be retained 
in the main belt region as scattering agents until most of the 
residual planetesimals are cleared. Due to dynamical friction, the 
velocity dispersion of the embryos is expected to be smaller than 
that of the lower-mass planetesimals \citep{wetherillstewart1989, 
palmer1993}.  However, a protracted retention of embryos may 
lead to the formation of terrestrial planets in the main belt
region. \cite{hansen2009} showed that the spacial distribution 
of the terrestrial planets requires their building blocks to 
be confined within $\sim 2$ AU.  A confinement mechanism has 
been suggested by \cite{walsh2011}.  This ``grand-tack" 
scenario is based on the assumption that due to its tidal
interaction with the disk gas (in consort with Saturn), Jupiter may 
have migrated inward through the solar nebula and then outward over 
a few AU to its present-day location.  During this course
of this excursion, most of the planetary building blocks would
be swept into the inner solar system.  The probability of such an
excursion is uncertain. Since Jupiter migration is driven by its
tidal interaction with the disk, this process cannot occur after 
the depletion of the disk gas.  

In the embedded-embryos scenario \citep{obrien2007}, the time 
scale for $90\%$ planetesimal clearing by an assumed population of 
embedded Mars-size embryos (with a total mass of $5 M_\oplus$) 
is $\sim 10$~Myr, which is comparable to or longer than the observed 
gas depletion time of $\sim 3-5$ Myr \citep{hartmann1998}. The  
``grand-tack" scenario would make the embedded-embryos scenario 
obsolete if both embryos and planetesimals are appreciably cleared 
from the main belt region by a migrating Jupiter. The removal
efficiency is independent of planetesimals' mass and radius, and 
therefore it does not lead to a size selection for the retained 
progenitors of asteroids.  Partial reduction
of the embryo population would reduce the scattering centers and 
their efficiency in the planetesimals' inclination excitation.  

In our model, the rate of planetesimals' orbital decay
is mostly determined by the magnitude of eccentricity and its damping 
timescale.  In the unperturbed regions, the velocity dispersion of
small planetesimals is modest and that of large planetesimals is 
small.  But their 
eccentricity is excited above $e>0.5$ during the passage of SRs or MMRs 
(Fig.~\ref{fig:e_a_t_rp100}). With relatively small damping timescale
(Fig.~\ref{fig:r_t1_t2_td_i_f}), the large planetesimals (embryos) 
undergo orbital decay in 
consort with the inwardly sweeping SRs.  In contrast, the eccentricity
damping is less efficient for the intermediate-size planetesimals, and
they are preferentially retained with eccentricities (Fig.~\ref{fig:a_e_r}) 
comparable to that ($e \sim 0.2$) of most asteroids with radius larger 
than 50~km \citep{knezevic2003}. 

The time scales for $\nu_5$ to propagate to the present semimajor 
axes of Mars and Earth are $4T_{\rm dep}$ and $10T_{\rm dep}$, or $\sim 12$ 
and $30$ Myr, for the observationally inferred value $T_{\rm dep} 
\sim 3$ Myr.  The passage of the $\nu_5$ SR leads to eccentricity 
excitation, which promotes orbit crossing and physical collisions 
among the inwardly migrating embryos.  \cite{nagasawa2005} and 
\cite{thommes2008} proposed a ``dynamical shake-up" model to account 
for the formation time scale of terrestrial planets as inferred from 
cosmochemical data \citep{kleine2009} and their modest eccentricities.

\subsection{Comparison with previous models on the initial 
eccentricity and semimajor axes of the Jupiter-Saturn system} 

Throughout this paper, we have emphasized that efficient damping of  
excited eccentricity is essential for the clearing of planetesimals. 
\cite{obrien2007} have already shown that with nearly circular 
orbits, Jupiter and Saturn SRs cannot significantly excite the 
planetesimals' eccentricity over the observed gas depletion time 
scale of $3-5$ Myr.  In model 
$A_1$, we show that a significant fraction of Jupiter and Saturn 
present-day angular momentum deficit is required for their SRs
to excite the planetesimals' eccentricity to an adequate level 
for the depletion of the initial planetesimals in the main belt
region to the current asteroid population.  We also simulated
model $A_4$ with a reduced amount of angular momentum deficit for the 
Jupiter and Saturn system. A comparison of the results generated 
from models $A_1$ and $A_4$ (Fig. \ref{fig:r_df_ej}) confirms the 
expectation that the efficiency of eccentricity excitation 
and planetesimal clearing increases with $e_J$.

Several previous models have suggested that Saturn may have migrated
outward over some significant distances.  For example, the Nice model
has attributed the LHB to the passage of Saturn
through Jupiter's 2:1 MMR which is currently located 
at 8.18~AU.  In model $A_{6}-A_{9}$, we consider the possibility of a 
closer initial Jupiter-Saturn separation.
The results of these simulations show that both the MMRs and SSRs can
efficiently clear both small and large planetesimals while retaining 
intermediate-size planetesimals in the main belt region if, during the
epoch of disk depletion, Jupiter and Saturn were sufficiently 
close to their present-day location (models $A_1$ and $A_8$). If
Jupiter and Saturn were close to each other's 2:1 MMR (model $A_9$)
during the epoch of nebula depletion, $<10^{-4}$ in mass of the 
initial population of planetesimals may be retained.  This total
mass is less than that in the asteroid belt.  More mass may be
retained to accommodate a sufficiently larger reservoir of residual
planetesimals for the LHB if the nebula depletion 
time scale is substantially less than 1 Myr.
These results place some constraints but do not eliminate the 
possibility that Saturn may have undergone extensive outward migration.
After the nebula depletion, passage of Saturn through its 2:1 MMR with Jupiter, as envisioned in the Nice model, may 
cause the excitation of terrestrial planets' as well as asteroids'
eccentricity beyond their present-day values \citep{agnorlin2012}. 

If Saturn was initially located 
between Jupiter's 2:1 MNR and its present-day location (e.g., with a semimajor axis $\sim 10-15\%$ smaller than its value today), the sweeping 
path of $\nu_6$ would be stalled outside the inner boundary of the main 
belt region.  Models $A_6$ and $A_7$ show that a population of 
planetesimals of all sizes would be retained around the provisional 
location of the $\nu_6$ SR (Fig.~\ref{fig:v56_r_af}). When Saturn 
eventually migrates to its present-day location after the disk gas 
is depleted, the eccentricity of the residual planetesimals would be 
excited as the $\nu_6$ SR sweeps past them, but they would not undergo 
orbital decay without any eccentricity damping. Although the 
orbits of some highly eccentric planetesimals may be destablized by 
the terrestrial planets'  perturbation, a large fraction of the 
residual planetesimals of all sizes would be retained within the 
main belt region.

\subsection{Future investigations}

We mainly focus on the dynamical depletion of asteroids in the main belt. 
During this process, the transitional disk and the gas giants are 
responsible for the gravitational perturbations on the asteroids. 
In this paper, we assumed a global depletion factor for the solar nebula.
Photoevaporation and disk winds may lead to preferential locations 
for mass loss.  Under some circumstances, inside-out clearing of the 
disk may promote planetesimals' inclination excitation.  In addition, 
the disk also provides a damping force that affects the embedded 
planetesimals. In most models, we have not included the mutual interaction 
between planetesimals. Although this assumption is adequate for the 
determination of the planetesimals' eccentricity and semimajor axis 
evolution, this effect needs to be included (as in the modified default 
model) to enable us to account for the excitation of planetesimals' 
inclination by embedded embryos (Fig. \ref{fig:a_i_r}).  We have also 
neglected collisions between asteroids and their subsequent fragmentation.  
Quantitative estimates suggest that 
the collision time scale is $>4.6$ Gyr for asteroids larger than $\sim 50-100$ 
km such that the transitions of SFD power index in this size range have not
evolved significantly since the formation of the solar system.  Nevertheless, 
collisions and fragmentations may have significantly modified the small-size 
range of the observed SFD. Follow-up investigations on planetesimal-embryo
interaction and on fragmentation process are warranted.

Our model supports the SSR model as the process that lead to the asteroids' present-day SFD. Since all planetary systems originally form in circumstellar gas disks that evolve and ultimately dissipate, 
this model, as studied in this paper, can be applied to construct models that
account for the kinematic structure of a wide range of observed extrasolar
planetary systems.  Such analysis will provide useful clues and place
important constraints on protoplanets' birth environment and their 
post-formation dynamical evolution (Zheng, X.C. et al, in preparation).



\section*{Acknowledgments}

We wish to thank the anonymous reference for carefully reading the manuscript and providing useful feedback that helped to improve the paper.
This work was supported in part by a UC/Lab Fee grant.  We thank Shude Mao, Qingzhu Yin,  Makiko Nagasawa, Shigeru Ida, and Munan Gong for useful and stimulating conversations.   
M.B.N.K. was supported by the Peter and Patricia Gruber Foundation through the PPGF fellowship, by the Peking University One Hundred Talent Fund (985), and by the National Natural Science Foundation of China (grants 11010237, 11050110414, 11173004, and 11573004). 
This publication was made possible through the support of a grant from 
the John Templeton Foundation and National Astronomical Observatories of the
Chinese Academy of Sciences. 


\begin{thebibliography}{}


\bibitem[Aarseth(1993)]{aarseth1993} Aarseth, S. J., Lin, D. N. C., \& Palmer, P. L. 1993, ApJ, 403, 351

\bibitem[Aarseth(2003)]{aarseth2003} Aarseth, S. J. 2003, Gravitational 
N-Body Simulations. Cambridge Univ. Press, Cambridge 

\bibitem[Adachi et al.(1976)]{adachi1976} Adachi, I., Hayashi, C., \& 
Nakazawa, K. 1976, Prog. Theor. Phys., 56, 1756

\bibitem[Agnor \& Asphaug(2004)]{agnor2004} Agnor, C., Asphaug E. 2004, ApJ, 613, L157

\bibitem[Agnor \& Lin(2012)]{agnorlin2012} Agnor, C. B., Lin, D. N. C. 2012, ApJ, 745, 143

\bibitem[Artymowicz(1993)]{artymowicz1993} Artymowicz, P. 1993, ApJ, 
419, 166

\bibitem[Bottke et al.(2005)]{bottke2005} Bottke, W. F., Durda, D. D., 
Nesvorny, D., Jedicke, R., Morbidelli, A., Vokrouhlicky, D., Levison, 
H. 2005, Icarus, 175, 111

\bibitem[Brauer et al.(2008)]{brauer2008} Brauer, F., Henning, T., \& Dullemond, C. P. 2008, A\&A, 487, L1


\bibitem[Bryden et al.(1999)]{bryden1999} Bryden, G., Chen, X., Lin, D. N. C., Nelson, R. P., Papaloizou J. C. B. 1999, ApJ, 514, 344

\bibitem[Bryden et al.(2000)]{bryden2000} Bryden, G., R\'o\.zyczka, M., Lin, D. N. C., Bodenheimer, R. 2000, ApJ, 540, 1091

\bibitem[Chambers(2001)]{chambers2001} Chambers, J.E. 2001, Icarus 152, 205

\bibitem[Chambers(2008)]{chambers2008} Chambers, J. 2008, Icarus, 198, 256

\bibitem[Ciesla \& Hood (2002)] {ciesla2002} Ciesla, F. J., Hood, L. L. 
2002, Icarus, 158, 281

\bibitem[Connolly et al.(1998)] {connolly1998} Connolly, H. C., Jr., 
Love, S. G. 1998, Sci, 280, 62

\bibitem[Currie \& Sicilia-Aguilar(2011)]{currie2011} Currie, T. \& Sicilia-Aguilar, A. 2011, ApJ, 732, 24 


\bibitem[Cuzzi(1993)] {cuzzi1993} Cuzzi, J. N., Dobrovolskis, A. R., Champney, J. M. 1993, Icarus, 106, 102

\bibitem[Cuzzi at al.(2008)] {cuzzi2008} Cuzzi, J. N., Hogan, R. C., 
Shariff, K. 2008, ApJ, 687, 1432

\bibitem[DeMeo \& Carry(2014)]{demeo2014} DeMeo, F. E., Carry, B. 2014, Nature, 505, 629

\bibitem[Descamps \& Marchis(2008)]{descamps2008} Descamps, P., and Marchis, F.  2008,  Icarus, 193, 74

\bibitem[Desch \& Connolly(2002)] {desch2002} Desch, S. J., Connolly, 
H. C., Jr. 2002, Meteoritics and Planetary Science, 37, 183

\bibitem[Dobbs-Dixon et al.(2007)] {dobbs2007} Dobbs-Dixon, I., Li, S.-L. \& Lin, D. N. C. 2007, ApJ, 660, 791

\bibitem[Dohnanyi(1969)] {dohnanyi1969} Dohnanyi, J. S. 1969, JGR, 74, 2531

\bibitem[Fernandez \& Ip(1984)]{fernandez1984} Fernandez, W.-H., Ip, J.A. 1984, Icarus 58, 109

\bibitem[Garaud \& Lin(2004)] {garaudlin2004} Garaud, P., Lin, D. N. C. 2004, ApJ, 608, 1050

\bibitem[Garaud \& Lin(2007)] {garaudlin2007} Garaud, P., \& Lin, D. N. C. 2007, ApJ, 654, 606

\bibitem[Goldreich \& Ward(1973)] {goldreichward1973} Goldreich, P., Ward, W. R., 1973, ApJ, 183, 1051

\bibitem[Goldreich \& Tremaine(1980)] {goldreichtremaine1980} Goldreich, P., \& Tremaine, S. 1980, ApJ, 241, 425

\bibitem[Gomes et al.(2005)]{gomes2005} Gomes, R., Levison, H.F., Tsiganis, K., Morbidelli, A. 2005, Nature 435, 466

\bibitem[Haisch et al.(2001)] {haisch2001} Haisch, K. E., Jr., Lada, E. A., Lada, C. J. 2001, ApJ, 553, 153

\bibitem[Hansen(2009)]{hansen2009} Hansen, B. M. S. 2009, ApJ, 703, 1131

\bibitem[Hartmann(1998)] {hartmann1998} Hartmann, L. 1998, Accretion Processes in Star Formation. Cambridge Univ. Press, Cambridge

\bibitem[Hayashi(1981)] {hayashi1981} Hayashi, C. 1981, Progress of 
Theoretical Physics Supplement 70, 35-53

\bibitem[Heppenheimer(1980)] {heppenheimer1980} Heppenheimer, T. A. 
1980, Icarus, 41, 76

\bibitem[Ida \& Lin(1996)] {ida1996} Ida, S., \& Lin, D. N. C. 1996, AJ, 112, 1239

\bibitem[Ida et al.(2000)]{ida2000} Ida, S., Bryden, G., Lin, D. N. C., Tanaka, H. 2000, ApJ, 534, 428

\bibitem[Ida \& Lin(2004)] {idalin2004} Ida, S. \& Lin, D. N. C. 2004, ApJ, 604, 388

\bibitem[Ida et al.(2013)] {idalin2013} Ida, S., Lin, D. N. C., \& Nagasawa, M. 2013, ApJ, 775, 42


\bibitem[Johansen et al.(2007)] {johansen2007} Johansen, A., Oishi, 
J. S., Low, M.-M. M., Klahr, H., Henning, T., Youdin, A. 2007, Nature, 448, 1022

\bibitem[Kelley \& Gaffey(2000)]{kelley2000} Kelley, M. S., Gaffey, M. J. 2000, Icarus, 144, 27

\bibitem[Kretke \& Lin(2007)]{kretke2007} Kretke, K. A. \& Lin. D. N. C. 2007, ApJ, 664, L55

\bibitem[Kleine et al.(2009)] {kleine2009} Kleine, T., Touboul,M., Bourdon, B., et al. 2009, Geochim. Cosmochim. Acta, 73, 5150

\bibitem[Knezevic \& Milani (2003)]{knezevic2003} Knezevic, Z., Milani, A. 2003, Astron. Astrophys. 403, 1165


\bibitem[Kokubo \& Ida(1998)] {kokuboida1998} Kokubo, E., Ida, S., 1998, Icarus, 131, 171


\bibitem[Lecar \& Franklin(1997)]{lecar1997} Lecar, M. \& Franklin, F. 1997, Icarus, 129, 134

\bibitem[Lee \& Peale(2002)]{lee2002} Lee, M. \& Peale, S. J. 2002, ApJ, 567, 596

\bibitem[Lemaitre \& Dubru(1991)]{lemaitre1991} Lemaitre, A. \& Dubru, P. 1991, CeMDA, 52, 57

\bibitem[Leinhardt \& Stewart(2012)] {leinhardt2012} Leinhardt, Z. M., Stewart, S. T. 2012, ApJ, 745, 79


\bibitem[Li et al.(2010)]{liagnorlin2010} Li, S. L., Agnor, C. B., \& Lin, D. N. C. 2010, ApJ, 720, 1161

\bibitem[Lin \& Papaloizou(1986)]{lin1986} Lin, D. N. C. \& Papaloizou, J. 1986, ApJ, 309, 846


\bibitem[Malhotra(1989)]{malhotra1989} Malhotra, R., Fox, K., Murray, C. D., \& Nicholson, P. D. 1989, A\&A, 221, 348

\bibitem[Malhotra(1993)]{malhotra1993} Malhotra, R. 1993, Nature 365, 819

\bibitem[Margot et al.(2002)]{margot2002} Margot, J. L., Nolan, M. C., Benner, L. A. M., et al. 2002, Sci, 296, 1445

\bibitem[Marchis et al.(2006)]{marchis2006} Marchis, F., Hestroffer, D., Descamps, P. et al., 2006, Nature, 439, 565

\bibitem[McSween(1999)]{mcsween1999} McSween, Harry Y. (1999). Meteorites and their parent planets. Cambridge Univ. Press, Cambridge

\bibitem[Morbidelli et al.(2005)]{morbidelli2005} Morbidelli, A., Levison, H.F., Tsiganis, K., Gomes, R. 2005, Nature 435, 462

\bibitem[Morbidelli et al.(2009)]{morbidelli2009} Morbidelli, A, Bottke, 
W, Nesvorny, D, Levison, H. F. 2009, Icarus, 204, 558

\bibitem[Murray \& Dermott(1999)]{murray1999} Murray, C. D., Dermott, D. F. 1999, Solar system Dynamics. Cambridge Univ. Press, Cambrige

\bibitem[Nagasawa \& Ida(2000)]{nagasawaida2000} Nagasawa, M., \& Ida, S. 
2000, AJ, 120, 3311

\bibitem[Nagasawa et al.(2000)]{nagasawa2000} Nagasawa, M., Tanaka, H., Ida, S. 2000, AJ, 119, 1480

\bibitem[Nagasawa et al.(2001)]{nagasawa2001} Nagasawa, M,. Ida, S., Tanaka, H. 2001, EP\&S, 53, 1085  

\bibitem[Nagasawa et al.(2002)]{nagasawa2002} Nagasawa, M., Ida, S., Tanaka, H. 2002, Icarus, 159, 322

\bibitem[Nagasawa et al.(2005)]{nagasawa2005} Nagasawa, M., Lin D., 
Thommes, E. 2005, ApJ, 635, 578


\bibitem[Nesvorny et al.(2006)]{nesvorny2006} Nesvorny, D., Vokrouhlicky, D., Bottke, W. F. 2006, Sci, 312, 1490


\bibitem[O'Brien et al.(2007)]{obrien2007} O'Brien, D. P., Morbidelli, A., Bottke, W. F. 2007, Icarus, 191,434

\bibitem[Paardekooper et al.(2011)]{paardekooper2011} Paardekooper, S. J., Baruteau, C., Kley W. 2011, MNRAS, 410, 293

\bibitem[Palmer et al.(1993)]{palmer1993} Palmer, P. L., Lin, D. N. C., \& Aarseth, S. J. 1993, ApJ, 403, 336

\bibitem[Petit et al.(2001)]{petit2001} Petit, J., Morbidelli, A., Chambers, J. 2001, Icarus 153, 338

\bibitem[Pollack et al.(1996)]{pollack1996} Pollack, J. B., Hubickyj, O., Bodenheimer, P., Lissauer, J. J., Podolak, M., \& Greenzweig, Y. 1996, Icarus, 124, 62


\bibitem[Safronov(1969)]{safronov1969} Safronov, V.S. (1969), Evolution of the Protoplanetary Cloud and Formation of the Earth and Planets, Nauka, Moscow. NASA TTF-677

\bibitem[Stewart \& Leinhardt(2012)]{stewart2012} Stewart, S. T., Leinhardt, Z. M. 2012, ApJ, 751, 32

\bibitem[Su et al.(2013)]{su2013}Su, K. Y. L., Rieke, G. H., Malhotra, R., et al. 2013, ApJ, 763, 118

\bibitem[Supulver \& Lin(2000)]{supulver2000} Supulver, KD, Lin D. N. C. 
2000, Icarus, 146, 525


\bibitem[Thommes et al.(2008)]{thommes2008} Thommes, E., Nagasawa, 
M., \& Lin, D. N. C. 2008, ApJ, 676, 728

\bibitem[Tsiganis et al.(2005)]{tsiganis2005} Tsiganis, K., Gomes, R., Morbidelli, A., Levison, H.F. 2005, Nature 435, 459

\bibitem[Walsh et al.(2011)]{walsh2011} Walsh, K. J., Morbidelli, A., Raymond, S. N., O'Brien, D. P., Mandell, A. M. 2011, Sci, 475, 206

\bibitem[Ward(1981)]{ward1981} Ward, W. R. 1981, Icarus, 47, 234

\bibitem[Ward(1989)]{ward1989} Ward, W. R. 1989, ApJ, 345, L99

\bibitem[Ward(1993)]{ward1993} Ward, W. R. 1993, Icarus, 106, 274 

\bibitem[Ward(1997)]{ward1997} Ward, W. R. 1997, Icarus, 126, 261

\bibitem[Weidenschilling(1977)]{weidenschilling1977} Weidenschilling, S. J. 1977, MNRAS, 180, 57

\bibitem[Weidenschilling \& Cuzzi(1993)]{weidenschillingcuzzi1993} Weidenschilling S. J., Cuzzi J. N., 1993, in Levy E. H., Lunine J. I., eds, Protostars and Planets III Formation of
planetesimals in the solar nebula. Protostars and Planets III, University of Arizona Press,
p.1031-1060

\bibitem[Weidenschilling et al.(1997)]{weidenschilling1997} Weidenschilling, S. J., Spaute, D., Davis, D. R., Marzari, F., Ohtsuki, K. 1997, Icarus, 128, 429


\bibitem[Wetherill(1980)]{wetherill1980} Wetherill, G. W. 1980, ARA\&A, 18, 77

\bibitem[Wetherill(1989)]{wetherill1989} Wetherill, George W., " Origin of the asteroid belt"
Asteroids II; Proceedings of the Conference, Tucson, AZ, Mar. 8-11, 1988 (A90-27001 10-91). Tucson, AZ, University of Arizona Press, 1989, p. 661-680

\bibitem[Wetherill \& Stewart(1989)]{wetherillstewart1989} Wetherill, G.W., Stewart, G.R. 1989, Icarus 77, 330

\bibitem[Wetherill(1992)]{wetherill1992} Wetherill, G.W. 1992, Icarus 100, 307

\bibitem[Whipple(1972)]{whipple1972} Whipple, F. L. 1972, in From Plasma to Planet, ed. A. Elvius (New York: Wiley), 211

\bibitem[Youdin \& Shu (2002)]{youdinshu2002} Youdin, A. N., Shu, F. H. 2002, ApJ, 580, 494

\bibitem[Youdin \& Goodman (2005)]{youdingoodman2005} Youdin, A., Goodman, J. 2005, ApJ, 620, 459

\bibitem[Zappala et al.(2002)]{zappala2002} Zappalà, V.; Cellino, A.; dell'Oro, A.; Paolicchi, P., "Physical and Dynamical Properties of Asteroid Families" in Asteroids III, W. F. Bottke Jr., A. Cellino, P. Paolicchi, and R. P. Binzel (eds), University of Arizona Press, Tucson, p.619-631

\bibitem[Zhou et al.(2007)]{zhou2007} Zhou, J. L., Lin, D. N. C., \& 
Sun, Y. S. 2007, ApJ, 666, 423

\bibitem[Zuckerman et al.(1995)]{zuckerman1995} Zuckerman, B., Forveille, F.,  Kastner, J. H. 1995, Nature, 373, 494 

\end{thebibliography}

\end{CJK*}
\end{document}